\documentclass[aps,prb,twocolumn,amsmath,amssymb,superscriptaddress]{revtex4-2}
\usepackage{graphicx}
\usepackage{amssymb}
\usepackage{color}
\usepackage{amsmath}
\usepackage{float}
\usepackage{gensymb}
\usepackage{chemformula}
\usepackage{epstopdf}
\usepackage{hyperref}
\usepackage{makecell}
\usepackage{multirow}
\usepackage{afterpage}
\usepackage{tabularx}
\usepackage{amsmath}
\usepackage{float}
\usepackage{booktabs}

\newcommand{\beq}{\begin{eqnarray}}
	\newcommand{\eeq}{\end{eqnarray}}

\begin{document}
	\title{Uniaxial strain tuned magnetism of the altermagnet candidate h-FeS}
	\author{Weiliang~Yao}
	\email{wy28@rice.edu}
	\affiliation{Department of Physics and Astronomy, Rice University, Houston, Texas 77005, USA}
	\affiliation{Rice Laboratory for Emergent Magnetic Materials and Smalley-Curl Institute, Rice University, Houston, Texas 77005, USA}
	\author{Feng~Ye}
	\affiliation{Neutron Scattering Division, Oak Ridge National Laboratory, Oak Ridge, Tennessee 37831, USA}
	\author{Zachary~J.~Morgan}
	\affiliation{Neutron Scattering Division, Oak Ridge National Laboratory, Oak Ridge, Tennessee 37831, USA}
	\author{Douglas~L.~Abernathy}
	\affiliation{Neutron Scattering Division, Oak Ridge National Laboratory, Oak Ridge, Tennessee 37831, USA}
	\author{Ruixian~Liu}
	\affiliation{Center for Advanced Quantum Studies and Department of Physics, Beijing Normal University, Beijing 100875, People's Republic of China}
	\author{Sijie~Xu}
	\affiliation{Department of Physics and Astronomy, Rice University, Houston, Texas 77005, USA}
	\affiliation{Rice Laboratory for Emergent Magnetic Materials and Smalley-Curl Institute, Rice University, Houston, Texas 77005, USA}
	\author{Yuxiang~Gao}
	\affiliation{Department of Physics and Astronomy, Rice University, Houston, Texas 77005, USA}
	\affiliation{Rice Laboratory for Emergent Magnetic Materials and Smalley-Curl Institute, Rice University, Houston, Texas 77005, USA}
	\author{Kevin~Allen}
	\affiliation{Department of Physics and Astronomy, Rice University, Houston, Texas 77005, USA}
	\affiliation{Rice Laboratory for Emergent Magnetic Materials and Smalley-Curl Institute, Rice University, Houston, Texas 77005, USA}
	\author{Yuan~Fang}
	\affiliation{Department of Physics and Astronomy, Rice University, Houston, Texas 77005, USA}
	\affiliation{Rice Laboratory for Emergent Magnetic Materials and Smalley-Curl Institute, Rice University, Houston, Texas 77005, USA}
	\author{Emilia~Morosan}
	\affiliation{Department of Physics and Astronomy, Rice University, Houston, Texas 77005, USA}
	\affiliation{Rice Laboratory for Emergent Magnetic Materials and Smalley-Curl Institute, Rice University, Houston, Texas 77005, USA}
	\author{Qimiao~Si}
	\affiliation{Department of Physics and Astronomy, Rice University, Houston, Texas 77005, USA}
	\affiliation{Rice Laboratory for Emergent Magnetic Materials and Smalley-Curl Institute, Rice University, Houston, Texas 77005, USA}
	\author{Pengcheng~Dai}
	\email{pdai@rice.edu}
	\affiliation{Department of Physics and Astronomy, Rice University, Houston, Texas 77005, USA}
	\affiliation{Rice Laboratory for Emergent Magnetic Materials and Smalley-Curl Institute, Rice University, Houston, Texas 77005, USA}
	
	\date{\today}
	
	\begin{abstract}
		Altermagnets are collinear magnetic materials with ‘alter’nating local crystalline environments, characterized by joint spin and crystalline symmetries that enable ferromagnetic-like transport properties but with vanishing net magnetization. Hexagonal FeS (h-FeS) is a recently identified altermagnet candidate that shows a spontaneous anomalous Hall effect (AHE) accompanied by a tiny net magnetization. Here, we show that both the spontaneous AHE and magnetization can be effectively suppressed by an in-plane compressive strain. Since neutron diffraction measurements show that the applied uniaxial strain only modifies the in-plane domain population but does not affect the in-plane magnetic structure, the major effect of the applied strain is to tune the small $c$-axis ferromagnetic moment. Our results demonstrate a strong correlation between the tiny net magnetization and the spontaneous AHE in h-FeS, and show that uniaxial strain provides an effective knob to tune both properties in this altermagnet candidate for spintronic applications.
	\end{abstract}
	
	\maketitle
	
	The identification of altermagnetism \cite{SmejkalPRX2022,SmejkalPRX2022_2} expands the classification of magnetic order beyond ferromagnetism and antiferromagnetism, establishing a third fundamental magnetic state since the discovery of antiferromagnetism in the 1930s \cite{Neel1936}. Altermagnetism is defined by a distinct symmetry class in which collinear magnetic order breaks time-reversal symmetry while preserving zero net magnetization \cite{SmejkalPRX2022,SmejkalPRX2022_2,BaiAFM2024,SongNatRevMat2025,FenderJACS2025}. This symmetry-based perspective has led to the reclassification of a number of previously known antiferromagnetic (AFM) materials as altermagnets \cite{SmejkalPRX2022_2,BaiAFM2024,ChenNature2025,BhattaraiPRM2025}, thereby stimulating renewed interest in re-examination of many long-studied magnetic materials.
	
	Similar to antiferromagnets, altermagnets possess no macroscopic magnetization, rendering them robust against external magnetic fields \cite{SmejkalPRX2022,SmejkalPRX2022_2,BaiAFM2024,SongNatRevMat2025,FenderJACS2025}. At the same time, their electronic band structures exhibit momentum-dependent spin splitting even in the absence of spin–orbit coupling (SOC), enabling spin-polarized currents analogous to those in ferromagnets \cite{SmejkalPRX2022,SmejkalPRX2022_2,BaiAFM2024,SongNatRevMat2025,FenderJACS2025}. As a result, altermagnets can host unusual transport phenomena, including a spontaneous anomalous Hall effect (AHE) \cite{SmejkalSciAdv2020}, which originates from an effective magnetic field associated with time-reversal symmetry breaking and alternating spin polarization, rather than from a net magnetization as in a simple ferromagnet.
	
	The spontaneous AHE is one of the key signatures of an altermagnet, its observation requires not only a sufficient density of charge carriers but also an appropriate magnetic structure \cite{SmejkalSciAdv2020,UrataPRM2024,YunpjQuantMat2025}. To date, the spontaneous AHE has been observed in only a handful of materials among the huge number of proposed altermagnet candidates \cite{WasscherSolidStateCommun1965,BetancourtPRL2023,KluczykPRB2024,ReichlovaNC2024,ZhouArxiv2026,ShaoArxiv2026}. This scarcity has greatly limited comprehensive studies of altermagnetism-induced spontaneous AHE and its response to external tuning parameters. In addition, the relationship between the spontaneous AHE and the detailed magnetic structures, such as spin canting, remains elusive.
	
\begin{figure}[t!]
		\hspace{0cm}\includegraphics[width=8.5cm]{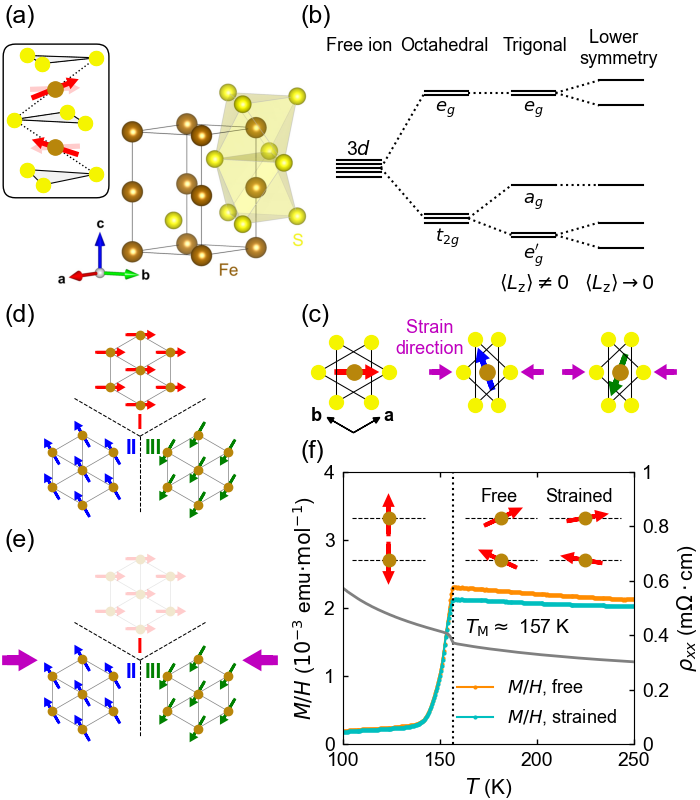}
		\caption{(a) Crystal structure of h-FeS above $T_{\rm{M}}$. Two S octahedra surrounding Fe ions are explicitly shown. The upper left inset shows the Fe spin configurations inside the two octahedra, where the spins initially along the [$1\bar{1}0$] direction exhibit a slight canting toward the Fe-S bonds (dashed lines). (b) Evolution of the orbital states for 3$d$ electrons under octahedral, trigonal, and lower-symmetry crystal fields. (c) Top view of the preferred spin directions in the FeS$_{6}$ octahedron in the free state and under compressive strain along the [$1\bar{1}0$] direction (magenta arrows). (d) Three magnetic domains (I, II, and III) in one Fe layer, connected by the $C_3$ symmetry. (e) Two remaining magnetic domains (II and III) under compressive strain along the [$1\bar{1}0$] direction. (f) Temperature dependence of the magnetic susceptibility $M/H$ (with and without strain, left axis) and the resistivity (right axis). A magnetic field of 1 T was applied along the $c$-axis for the $M/H$ data. The inset schematically shows a side-view of the spin configurations. Dashed vertical line marks $T_{\rm{M}}$. The canting angles are intentionally enlarged for better visualization.}
		\label{fig1}
	\end{figure}

	The hexagonal phase of FeS (h-FeS) has recently been identified as an altermagnet candidate showing an intriguing spontaneous AHE \cite{TakagiNM2025}. Due to the slight Fe deficiency, the exact crystal structure of h-FeS at a specific temperature depends on the stoichiometry of Fe and S \cite{HorwoodJSolidStateChem1976,Coey1976,WangPhaseTransition2005,TakagiNM2025}. For our samples, the actual chemical composition is close to Fe$_{0.93}$S \cite{SM}. At room temperature, these samples exhibit a NiAs-type structure (space group $P$6$_3$/$mmc$) [Fig. \ref{fig1}(a)] \cite{SM}, where the Fe ions form a triangular lattice in the $ab$-plane. Similar to the $g$-wave altermagnets $\alpha$-MnTe \cite{LeePRL2024,KrempaskyNature2024} and CrSb \cite{ReimersNC2024,DingPRL2024}, the altermagnetism of h-FeS originates from the alternating arrangement of the S octahedra surrounding the Fe ions [Fig. \ref{fig1}(a)]. Such structure preserves an overall six-fold rotational ($C_6$) symmetry even though each individual FeS$_6$ octahedron possesses only a local three-fold rotational ($C_3$) symmetry. Under the octahedral crystal field, the Fe$^{2+}$ ions adopt a high-spin electronic configuration $t_{\rm{2g}}^{4}e_{\rm{g}}^{2}$ \cite{SakkopoulosJPhysChemSolids1984,BansalNP2020}, with the orbital states further split by the trigonal and lower-symmetry crystal fields [Fig. \ref{fig1}(b)].
	
	Below $T_{\rm N} \approx 600$ K, the localized Fe moments develop a long-range magnetic order, aligning predominantly parallel within the $ab$-plane and antiparallel between adjacent triangular layers [Fig. \ref{fig1}(a)] \cite{HorwoodJSolidStateChem1976,Coey1976,WangPhaseTransition2005,BansalNP2020}. This so-called $A$-type AFM structure is further perturbed by additional local magnetocrystalline anisotropy, which induces a small spin canting and gives rise to a tiny net magnetization along the $c$-axis on the order of 10$^{-3}$ $\mu{\rm B}$/Fe \cite{TakagiNM2025}. Owing to the lattice’s $C_6$ symmetry, three equivalent in-plane magnetic domains are present [domains I, II, and III in Fig. \ref{fig1}(d)]. The spontaneous AHE was observed in this easy-plane spin configuration, which is consistent with symmetry analysis and regarded as direct evidence of the underlying altermagnetism \cite{TakagiNM2025}. At lower temperatures, the Fe moments reorient along the $c$-axis through a Morin transition \cite{VanDenBerg1972,HorwoodJSolidStateChem1976,Coey1976,WangPhaseTransition2005,BansalNP2020}, where the spontaneous AHE is symmetry-forbidden. Although the Morin transition temperature ($T_{\rm{M}}$) depends on the Fe and S stoichiometry, the main magnetic structures above and below it are insensitive to the precise value of $T_{\rm{M}}$ \cite{Andresen1967,VanDenBerg1972,HorwoodJSolidStateChem1976,Coey1976,WangPhaseTransition2005,BansalNP2020}.
	
	\begin{figure}[t!]
		\hspace{0cm}\includegraphics[width=8.8cm]{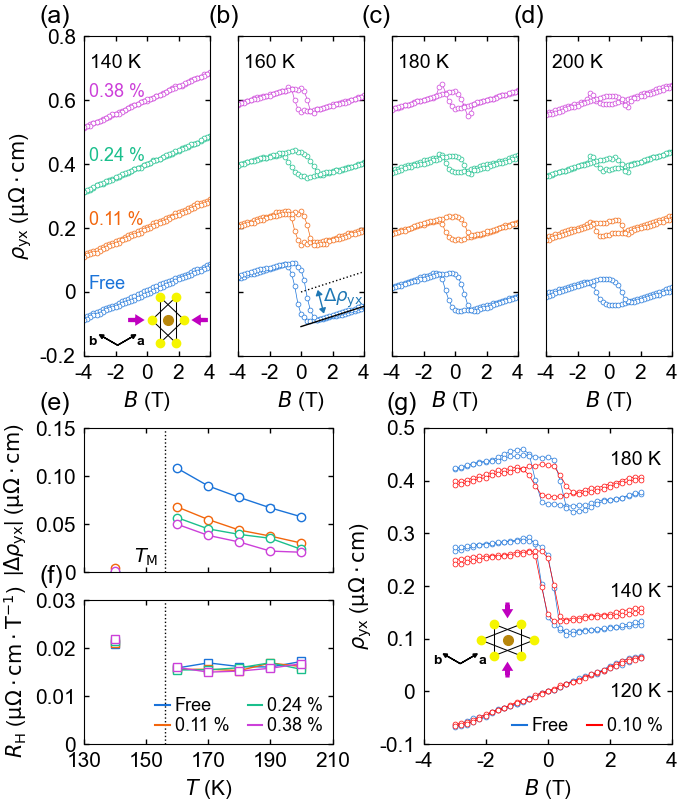}
		\caption{(a)-(d) Magnetic-field dependence of the Hall resistivity measured at selected temperatures under compressive strain applied along [$1\bar{1}0$]. The solid line in (b) represents a linear fit to the data, as described in the text. For clarity, the data are vertically offset by 0.2 $\mu \Omega\cdot\rm{cm}$. (e) and (f) Temperature dependence of the fitted anomalous Hall resistivity and the $B$-linear Hall coefficient at different strain magnitudes. (g) Magnetic-field dependence of the Hall resistivity measured at selected temperatures under compressive strain applied along [110]. Insets in (a) and (g) illustrate the corresponding strain directions.}
		\label{fig2}
	\end{figure}
	
	Strain has been widely thought to be an effective tool for controlling altermagnetism, ranging from modifying electronic band structures \cite{DevarajPRM2024,BelashchenkoPRL2025,TakahashiPRB2025,KarettaPRB2025,ChenArxiv2026} to tuning spin configurations \cite{AoyamaPRM2024,ZhouNature2025,KomuroPRB2025}. In this work, we explore the impact of uniaxial strain on the magnetic properties of h-FeS. We find that both the spontaneous AHE and the small net magnetization are simultaneously suppressed by compressive strain applied within the $ab$-plane. Neutron diffraction measurements under uniaxial strain further show that while the overall magnetic structure remains unchanged, the reduction of the AHE and magnetization is accompanied by a repopulation of in-plane magnetic domains. Such facile tunability of magnetic domains is consistent with a vanishingly small spin-wave gap, as directly revealed by inelastic neutron scattering (INS). These results demonstrate that the strain-reduced spontaneous AHE in h-FeS is correlated with suppressed spin canting, both of which are rooted in SOC and associated magnetoelastic coupling.
	
	Fig. \ref{fig1}(f) shows the magnetic susceptibility measured at 1 T as a function of temperature, which exhibits a sudden drop below $T_{\rm{M}}$. This behavior indicates a transition of the magnetic structure from an easy-plane spin configuration at high temperatures to an easy-axis spin configuration at low temperatures [see the inset of Fig. \ref{fig1}(f)]. Consistent with this magnetic transition, the temperature dependence of the in-plane resistivity displays a clear kink at $T_{\rm{M}}$, reflecting the influence of the magnetic structure change on electrical transport. These characterization results are qualitatively consistent with previous reports \cite{Coey1976,HorwoodJSolidStateChem1976,TakagiNM2025}, although the exact value of $T_{\rm{M}}$ depends on the chemical composition.
	
	Fig. \ref{fig2}(a)-(d) present the Hall resistivity at selected four temperatures. At 140 K, which is below $T_{\rm{M}} \approx$ 156 K for this sample, the Hall resistivity varies linearly with magnetic field, indicating that the response is dominated by the ordinary Hall effect (proportional to the magnetic field) and the conventional AHE (proportional to the magnetization). A spontaneous AHE emerges above $T_{\rm{M}}$ when the spins lie in the $ab$-plane, and its magnitude decreases with increasing temperature. Upon applying a compressive strain along [$1\bar{1}0$] direction \cite{note1}, the spontaneous AHE is gradually suppressed [Fig. \ref{fig2}(b)-(d)], while the contributions from ordinary Hall effect and conventional AHE are basically unchanged.
	
	To quantitatively determine the spontaneous AHE, we fit the high-field region ($B > 2$ T) of the Hall resistivity using the phenomenological expression $\rho_{yx} (B) = R_{\rm{H}} B + \Delta \rho_{yx}$, where the first term represents the $B$-linear contribution from the ordinary Hall effect and the conventional AHE (excluding the spontaneous part), and the second term corresponds to the spontaneous AHE. A representative fit for the data at 160 K without applied strain is shown in Fig. \ref{fig2}(b). The temperature dependencies of the fitting parameters ($R_{\rm{H}}$ and $\rho_{yx}$) under various strain conditions are summarized in Fig. \ref{fig2}(e) and (f), respectively. Assuming $\Delta \rho_{yx}$ is a signature of altermagnetism in h-FeS \cite{TakagiNM2025}, its systematic suppression suggests that altermagnetism can be effectively tuned by an in-plane strain. In contrast, the strain dependence of $R_{\rm{H}}$ [Fig. \ref{fig2}(f)] suggests that the overall electronic band structure is not dramatically affected by the size of the applied strain. We further note that the spontaneous AHE is similarly suppressed by compressive strain applied along [110] direction (perpendicular to [$1\bar{1}0$]), as shown in Fig. \ref{fig2}(g).
	
\begin{figure}[t!]
		\hspace{0cm}\includegraphics[width=8.5cm]{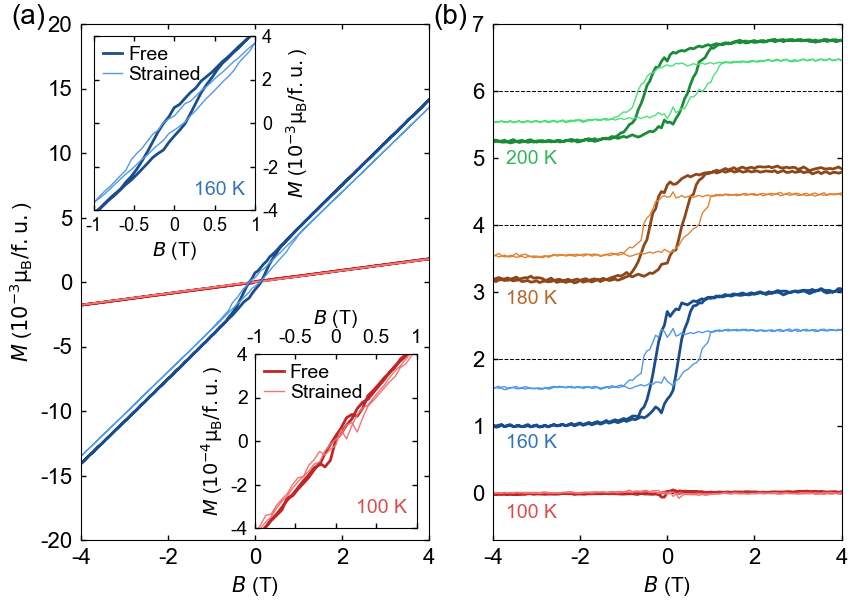}
		\caption{(a) Isothermal magnetization along the $c$-axis with and without compressive strain applied along [$1\bar{1}0$] at 100 K ($< T_{\rm{M}}$) and 160 K ($> T_{\rm{M}}$). Insets show the zoom-in view of data within $\pm$ 1 T. (b) Isothermal magnetization along the $c$-axis at selected temperatures after subtracting the high-field linear part. Light and dark colors denote data with and without strain, respectively. Data are evenly offset 0.002 $\mu_{\rm{B}}$/f.u. for clarity.}
		\label{fig3}
	\end{figure}

	Having established the strain-tuned spontaneous AHE, we next investigate the effect of uniaxial strain on the magnetization. Fig. \ref{fig3}(a) shows the isothermal magnetization measured along the $c$-axis direction. Below $T_{\rm{M}}$, the magnetization arises solely from the AFM structure and thus exhibits a linear dependence on the applied magnetic field. When spin canting develops above $T_{\rm{M}}$, a small hysteresis loop emerges in the low-field region, indicating a weak ferromagnetic contribution, although the overall magnetization remains predominantly linear with respect to the field. Therefore, above $T_{\rm{M}}$, the magnetization consists of both an AFM component and a ferromagnetic component. With the application of a compressive strain along [$1\bar{1}0$] direction, the small ferromagnetic component that appears only above $T_{\rm{M}}$ is reduced [Fig. \ref{fig3}(a)]. However, the differential AFM susceptibility (i.e., the slope of the linear background) is not affected both above and below $T_{\rm{M}}$. This behavior is further corroborated by the temperature dependence of the magnetic susceptibility shown in Fig. \ref{fig1}(f), where $M/H$ is reduced by strain above $T_{\rm{M}}$, but remains nearly unchanged below $T_{\rm{M}}$. To isolate the strain effect, we present the magnetization data after subtracting the AFM contribution, as shown in Fig. \ref{fig3}(b). It can be more clearly seen that the magnitude of the ferromagnetic component is suppressed under strain.
	
	Since the ferromagnetic component originates from spin canting \cite{TakagiNM2025}, these results indicate that compressive strain applied within the Fe triangular plane [Fig. \ref{fig1}(d) and (e)] effectively suppresses the $c$-axis spin canting in h-FeS. On the same sample used for these magnetization measurements, we have independently confirmed the strain-suppressed spontaneous AHE \cite{SM}. Although the small ferromagnetic component is believed to be noncritical for the spontaneous AHE \cite{TakagiNM2025}, the concurrent suppression of both the AHE and the ferromagnetic moment under in-plane uniaxial strain suggests a close correlation between the two phenomena in h-FeS. On the other hand, the field-induced AFM magnetization does not contribute the spontaneous AHE. Based on the level of strain we applied, estimated to be approximately 100 MPa, we determine the piezomagnetic coefficient of h-FeS to be $\sim$10$^{-3}$ $\rm{G} \cdot \rm{MPa}^{-1}$ at 160 K. This value is about one order of magnitude smaller than in Mn$_3$Sn \cite{IkhlasNP2022}, yet three orders of magnitude larger than in $\alpha$-MnTe \cite{AoyamaPRM2024}, both of which, interestingly, also exhibit strain-tuned spontaneous AHEs \cite{IkhlasNP2022,LiuArxiv2025,SmolenskiArxiv2025}.
	
	\begin{figure*}[t!]
		\centering{\includegraphics[width=0.92\textwidth]{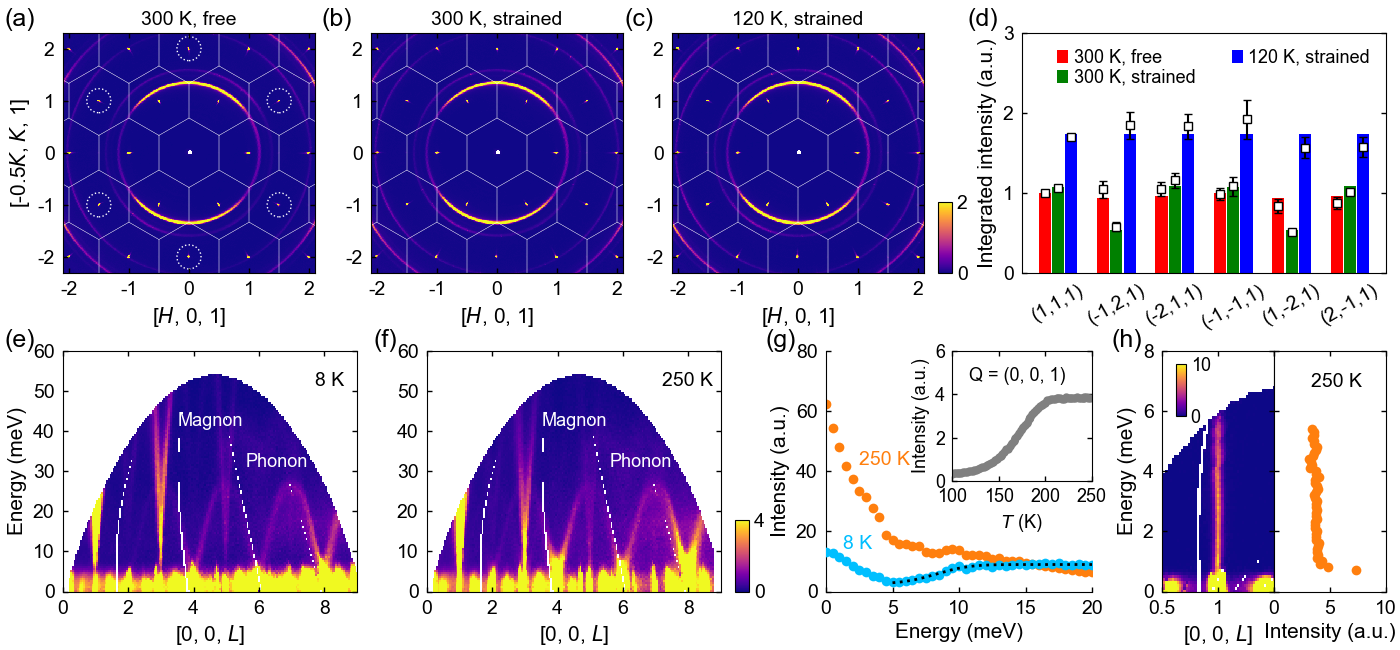}} \caption{(a)-(c) Neutron diffraction patterns in the ($H$, $K$, 1) plane measured at 300 K without strain, 300 K with strain, and 120 K with strain, respectively. White hexagons indicate the Brillouin zone boundaries. Dashed circles in (a) mark the Bragg peaks whose integrated intensities are shown in (d). Compressive strain is applied along the horizontal axis in (b) and (c). (d) Integrated intensities (square symbols with error bars) of six Bragg peaks under the three conditions. The bar chart represents the corresponding calculated intensities \cite{SM}. (e) and (f) INS spectra measured along the [0, 0, $L$] direction at 8 K and 250 K, respectively. (g) Energy dependence of the excitation intensity at (0, 0, 1) measured at 8 K and 250 K. The dashed curve shows a fit to the spin-wave gap using an error function \cite{SM}. The inset displays the temperature dependence of the diffraction intensity at the (0, 0, 1) Bragg peak. (h) Left: low-energy excitation spectrum around (0, 0, 1) at 250 K. Right: the corresponding energy dependence of the intensity at (0, 0, 1).}
		\label{fig4}
	\end{figure*}
	
	To gain further insight into the effects of in-plane compressive strain, we performed neutron diffraction measurements on h-FeS single crystals. Fig. \ref{fig4}(a), (b) and (c) show neutron diffraction patterns in the ($H$, $K$, 1) plane under three conditions: 300 K without applied strain, 300 K with compressive strain along the [$1\bar{1}0$] direction (above $T_{\rm{M}}$), and 120 K with compressive strain along the same direction (below $T_{\rm{M}}$). Although all Bragg peaks remain at integer-index positions, the intensities of some peaks exhibit pronounced changes. Fig. \ref{fig4}(d) shows the integrated intensities of the (1, 1, 1) Bragg peak and its five symmetry-equivalent counterparts in the unstrained case [Fig. \ref{fig4}(a)]. Because the nuclear structure factor vanishes at these reflections, their intensities can be attributed solely to the magnetic scattering. At 300 K without strain, the six peaks show nearly identical intensities. However, when the strain is applied, the intensities of ($-1$, $2$, $1$) and ($1$, $-2$, $1$) peaks are significantly reduced, whereas those of the remaining four peaks increase slightly. Such behavior can be well accounted for by strain-induced redistribution of in-plane magnetic domains above $T_{\rm{M}}$ \cite{SM}, and closely resembles that observed in $\alpha$-MnTe \cite{LiuArxiv2025}, which hosts the same $A$-type AFM structure. In the absence of strain, the three magnetic domains are equally populated [Fig. \ref{fig1}(d)]. Upon applying strain, however, the magnetic domain whose spins are aligned along the strain direction [i.e., domain I in Fig. \ref{fig1}(d)] is strongly suppressed, leaving only the other two domains with dominant populations [Fig. \ref{fig1}(e)].
	
	From the measured integrated intensities, we estimate that the population ratio of domain I : domain II : domain III is 0.31(4) : 0.33(4) : 0.36(4) in the absence of strain, and changes to 0.02(4) : 0.49(3) : 0.49(3) under applied strain \cite{SM}. Below $T_{\rm M}$, the intensities of the six peaks become comparable again even in the presence of strain [Fig. \ref{fig4}(d)], consistent with the easy-axis spin configuration. This behavior further confirms that the observed intensity variations originate from the magnetic response to strain, rather than from changes in the lattice itself. We point out that the intensity change due to strain-induced spin canting is on the order of 10$^{-7}$ \cite{SM}, as estimated from the $c$-axis ferromagnetic moment in the magnetization measurements (Fig. \ref{fig3}), and therefore lies well below the sensitivity of the unpolarized neutron diffraction technique used here. Moreover, such in-plane strain-driven domain repopulation is not expected to affect the strength of the spontaneous AHE, which behaves as an axial vector oriented along the $c$-axis. A detailed symmetry analysis of the conductivity tensor under strain is provided in \cite{SM}.
	
	The ease with which magnetic domains can be tuned by strain is also manifested in the spin excitations. Fig. \ref{fig4}(e) and (f) present the INS-measured excitation spectra along the [0, 0, $L$] direction below and above $T_{\rm M}$, respectively, where dispersive spin waves emanating from $L$= 1 and 3 are observed. From the energy dependence of the intensity at (0, 0, 1) [Fig. \ref{fig4}(g)], we find that the spin-wave gap of 8.3(3) meV at 8 K is filled in at 250 K, accompanied by an accumulation of elastic magnetic scattering intensity at (0, 0, 1) [inset of Fig. \ref{fig4}(g)]. With improved energy resolution, the spin-wave gap above $T_{\rm M}$ is found to be smaller than 0.5 meV, as shown in Fig. \ref{fig4}(g), which corresponds to the energy resolution at zero energy transfer \cite{SM}. The strongly reduced gap indicates that spins can more readily overcome the energy barrier separating the three magnetic domains in the easy-plane spin configuration. Consequently, the application of strain, even on the order of 100 MPa, is sufficient to lift the energy degeneracy among the three in-plane spin directions and selectively stabilize two domains. This situation is very similar to spin-wave gaps in pure and Li-doped $\alpha$-MnTe, where Li-doping into pure $\alpha$-MnTe induces a large spin anisotropy gap of $\sim$6 meV and rotate the moment from the in-plane to the $c$-axis direction \cite{SzuszkiewiczPRB2006,LiuPRL2024,YumnamPRB2024,ZhangArxiv2025}.
	
	Finally, we discuss the origin of the strain-tuned spin canting in terms of subtle strain-induced changes in the local coordination of the Fe ions. They are closely tied to magnetocrystalline anisotropy of h-FeS, which has been extensively studied for over six decades \cite{AdachiJPSJ1961,Adachi1963,SatoJPSJ1966,AdachiJApplPhys1968,HorwoodJSolidStateChem1976}. We first note that the orientation of the Fe magnetic moments is collectively determined by the crystal field, exchange interactions, and SOC, all of which depend sensitively on the lattice parameters $a$ and $c$ within the NiAs-type structure  \cite{AdachiJPSJ1961,Adachi1963,SatoJPSJ1966,AdachiJApplPhys1968,HorwoodJSolidStateChem1976}. At high temperatures, where the ratio $c/a$ is smaller than a critical value, the crystal field and exchange interactions together favor an easy-plane anisotropy, while the SOC promotes an out-of-plane spin canting \cite{AdachiJPSJ1961,HorwoodJSolidStateChem1976,BickfordPR1960,SlonczewskiPR1958}. Since the former two interactions are much stronger than the SOC, the spins exhibit only a small canting angle while remaining predominantly in plane. When a uniaxial strain is applied, the local $C_3$ symmetry of the FeS$_6$ octahedron is broken [Fig. \ref{fig1}(c)]. The partially quenched orbital angular momentum is further reduced, thereby lifting the remaining orbital degeneracy \cite{Fazekas1999,Khomskii2014} [Fig. \ref{fig1}(b)]. This additional quenching of the orbital angular momentum naturally weakens the effectiveness of SOC \cite{BickfordPR1960,SlonczewskiPR1958,HamPR1965,ChappertPRB1970}, leading to a reduction of the spin canting and, consequently, a suppression of the spontaneous AHE. A more detailed analysis is provided in \cite{SM}. This behavior re-manifests the famous Jahn-Teller effect, where the lattice distortion spontaneously occurs accompanied by quenching of the orbital angular momentum \cite{OBrienAmJPhys1993}. We note that the established magnetocrystalline anisotropy theory has also successfully explained both the Mori transition at $T_{\rm M}$ and the reduction of spin canting with increasing temperature (Fig. \ref{fig3}) \cite{AdachiJPSJ1961,Adachi1963,SatoJPSJ1966,AdachiJApplPhys1968}. This framework is again consistent with the experimentally observed suppression of the spontaneous AHE at elevated temperatures (Fig. \ref{fig2}) \cite{TakagiNM2025}.
	
	The direct correlation between the spontaneous AHE and the small ferromagnetic moment indicates that a finite magnetization plays a vital role in shaping the key observable of altermagnetism. In conventional ferromagnets, the spontaneous anomalous Hall resistivity ($\rho^{\rm A}$) is proportional to the net magnetization ($M$), $\rho^{\rm A} = R_{\rm S}\mu_0 M$, where $R_{\rm S}$ is the anomalous Hall coefficient and $\mu_0$ is the vacuum permeability \cite{NagaosaRMP2010}. Our results therefore suggest that h-FeS in essence behaves like a ferromagnet with a small net magnetization but an unusually large $|R_{\rm S}|$ (on the order of 100 $\mu\Omega\cdot\mathrm{cm}\cdot\mathrm{T}^{-1}$), which is significantly larger than in typical ferromagnets. For comparison, $|R_{\rm S}|$ is below 1 $\mu\Omega\cdot\mathrm{cm}\cdot\mathrm{T}^{-1}$ in elemental Fe and Ni \cite{Volkenshtein1960,KaulPRB1979}. Additional theoretical investigations are required to elucidate the microscopic origin of the large $|R_{\rm S}|$ in h-FeS.
	
	Notably, h-FeS belongs to a broader class of compensated magnets exhibiting large spontaneous AHEs, which include $\alpha$-MnTe \cite{WasscherSolidStateCommun1965,BetancourtPRL2023,KluczykPRB2024,ZhouArxiv2026,ShaoArxiv2026}, Mn$_3A$ ($A$ = Sn, Ge, and Pt) \cite{NakatsujiNature2015,AjayaSciAdv2016,LiuNatElectron2018}, Ce$_2B$Ge$_6$ ($B$ = Cu and Pd) \cite{KotegawaPRL2024}, and Co$C_3$S$_6$ ($C$ = Nb and Ta) \cite{GhimireNC2018,TakagiNP2023,ChenNanoLett2025}. Their spontaneous AHEs have been attributed either to nontrivial band structure effect (Berry curvature) in reciprocal space, or real space non-coplanar spins and associated nonzero scalar spin chirality. Nevertheless, all of these materials exhibit a small but detectable net magnetization concomitant with the AHE. While this magnetization is generally considered too small to account for the large spontaneous AHE, it would be interesting to experimentally examine whether simultaneous tuning of the magnetization and the spontaneous AHE, as established in h-FeS, can also be achieved in these materials.
	
	In summary, by combining electrical transport, magnetization, and neutron scattering measurements, we have systematically studied the magnetism of the altermagnet candidate h-FeS under in-plane compressive strain. We find that both the spontaneous AHE and the small ferromagnetic moment are suppressed by a uniaxial strain, which can be understood in terms of strain-induced tuning of the local coordination environment of the magnetic Fe ions. Neutron diffraction further reveals that, while the dominant magnetic structure remains unchanged, strain redistributes the magnetic domains of the easy-plane spin configuration, consistent with the negligible spin-wave gap observed by INS. Our results demonstrate that the spontaneous AHE in h-FeS, a hallmark of altermagnetism, is closely correlated with the weak ferromagnetic moment arising from spin canting, and establish uniaxial strain as an effective control parameter for tuning magnetism in altermagnet candidates.
	
	\begin{acknowledgments}
		The single-crystal synthesis, transport, magnetization, and neutron scattering experiments at Rice were supported by the U.S. DOE, BES under Grant Nos. DE-SC0012311 (P.D.) and DE-SC0026179 (P.D., Q.S. E.M.). Part of the materials characterization efforts at Rice is supported by the Robert A. Welch Foundation Grant No. C-1839 (P.D.). A portion of this research used resources at the Spallation Neutron Source, a DOE Office of Science User Facility operated by the Oak Ridge National Laboratory. The beam time was allocated to CORELLI on Proposal No. IPTS-34949.1 and ARCS on Proposal No. IPTS-34948.1.
	\end{acknowledgments}
	
	\bibliographystyle{apsrev4-1}
	\bibliography{FeS_strain_reference}
	
	\pagebreak
	\pagebreak

	\widetext
	
	\begin{center}
		\textbf{\large Supplemental Material for ``Uniaxial strain tuned magnetism of the altermagnet candidate h-FeS''}
	\end{center}
	
	\setcounter{equation}{0}
	\setcounter{table}{0}
	\setcounter{page}{1}
	\makeatletter
	\renewcommand{\theequation}{S\arabic{equation}}
	\renewcommand{\thefigure}{S\arabic{figure}}

\section{h-FeS single crystal growth method}
\begin{figure}[h!]
	\hspace{0cm}\includegraphics[width=8.5cm]{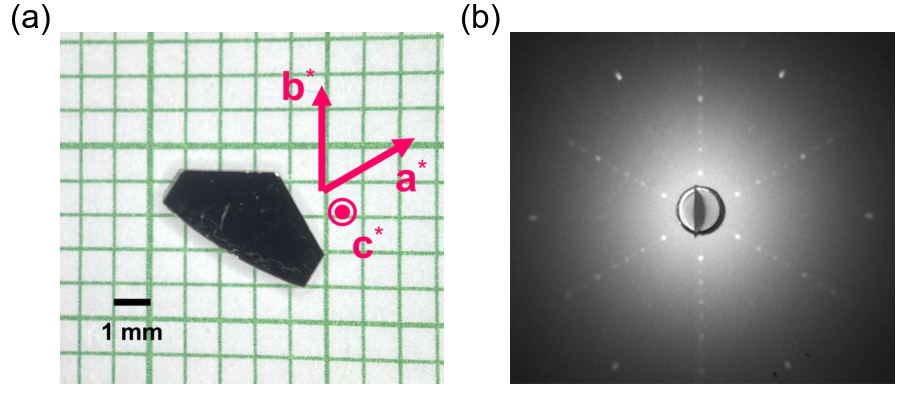}
	\caption{(a) One piece of representative h-FeS single crystal against a millimeter grid. (b) X-ray Laue diffraction pattern taken with the X-ray beam along the $c$-axis. The pattern corresponds to the sample orientation shown in (a).}
	\label{figs1}
\end{figure}

Hexagonal FeS (h-FeS) single crystals were grown using a chemical vapor transport method, slightly modified from previous reports \cite{KrabbesZAAC1975,KrabbesZAAC1976,KrabbesZAAC1976_2,Binnewies2017,BansalNP2020}. A mixture of 3.81 g iron powder, 2.28 g sulfur powder, and 0.22 g iodine was loaded into a quartz tube (outer diameter 28 mm, wall thickness 1.5 mm), which was then evacuated and sealed under vacuum. The sealed portion of the tube was approximately 23 cm in length. The tube was placed in a tube furnace, where the temperatures of the hot and cold ends were independently and sequentially controlled. The starting materials were positioned at the hot end of the tube. The hot and cold ends were heated to 940 $^{\circ}$C and 840 $^{\circ}$C, respectively, over 10 hours and maintained at these temperatures for 240 hours. Afterward, the furnace was turned off and allowed to cool naturally to room temperature. Single crystals with hexagonal edges [Fig. \ref{figs1}(a)] were obtained at the cold end of the tube. We find that using fine Fe powder is advantageous for growing larger crystals, although the physical properties of the obtained samples remain the same.

\section{Basic characterization of h-FeS single crystals}

Fig. \ref{figs1}(b) presents the X-ray Laue diffraction pattern taken with the X-ray beam along the $c$-axis, which indicates the six-fold rotational symmetry in the reciprocal space. The natural hexagonal edge is perpendicular to the $\mathbf{a}^{\ast}$ direction within the NiAs-type structure [Fig. \ref{figs1}(a)].

\begin{figure}[h!]
	\hspace{0cm}\includegraphics[width=13cm]{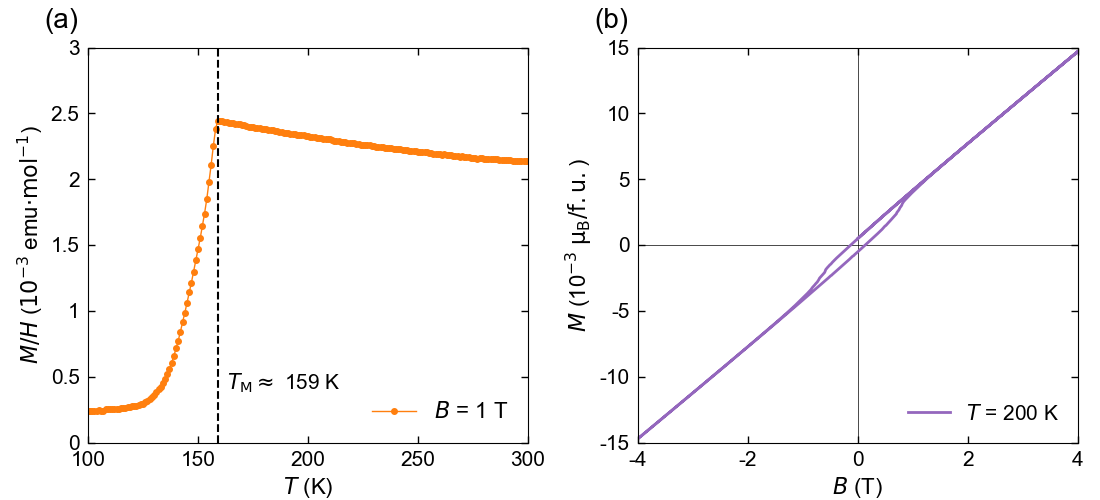}
	\caption{(a) Temperature dependence of the magnetic susceptibility under a magnetic field of 1 T applied along the $c$-axis. (b) Isothermal magnetization along the $c$-axis at 200 K.}
	\label{figs2}
\end{figure}

Fig. \ref{figs2}(a) shows the temperature dependence of the magnetic susceptibility of a piece of h-FeS single crystal. The measurement was taken with a Quantum Design PPMS system using the vibrating sample magnetometer (VSM), and a magnetic field of 1 T was applied along the $c$-axis. $M/H$ shows an abrupt drop below $\sim$159 K, indicating the occurrence of the Mori transition. The $c$-axis isothermal magnetization at 200 K ($> T_{\rm M}$) is shown in Fig. \ref{figs2}(b), where a hysteresis like the one in a ferromagnet can be observed in a small field region.

\begin{figure}[h!]
	\hspace{0cm}\includegraphics[width=11cm]{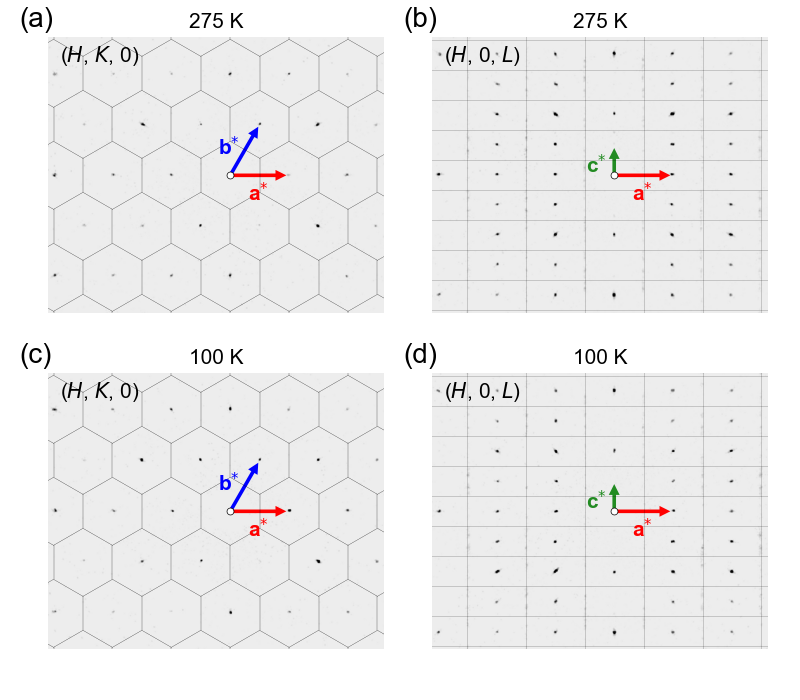}
	\caption{(a) and (b) Single crystal XRD patterns of ($H$, $K$, 0) and ($H$, 0, $L$) planes at 275 K. (c) and (d) Single crystal XRD patterns of ($H$, $K$, 0) and ($H$, 0, $L$) planes at 100 K. Hexagons and rectangles are the boundaries of the Brillouin zone.}
	\label{figs3}
\end{figure}

\begin{figure}[h!]
	\hspace{0cm}\includegraphics[width=12cm]{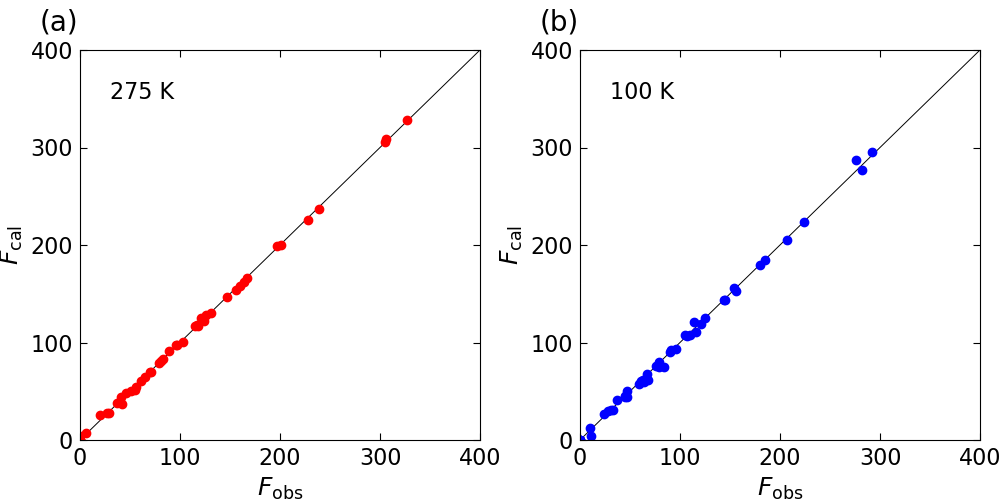}
	\caption{Calculated structure factor $F_{\rm{cal}}$ versus observed structure factor $F_{\rm{obs}}$ from single crystal XRD measured at 275 K (a) and 100 K (b).}
	\label{figs4}
\end{figure}

We crushed the exact single crystal used for the magnetic susceptibility measurements and selected a small fragment for single-crystal X-ray diffraction (XRD). The measurements were performed using a Rigaku SmartLab II X-ray diffractometer at 275 K (above $T_{\rm M}$) and 100 K (below $T_{\rm M}$). Representative diffraction patterns are shown in Fig. \ref{figs3}. Crystal structure refinements were carried out using the crystallographic program Jana2020 \cite{Petricek2023}. The calculated structure factors $F_{\rm cal}$ versus the observed structure factors $F_{\rm obs}$ at the two temperatures are presented in Fig. \ref{figs4}, with the corresponding refinement parameters summarized in Table \ref{tb1} and Table \ref{tb2}. The refined chemical composition is close to Fe$_{0.93}$S, which corresponds to a $T_{\rm M}$ of approximately 150 K based on the established relationship between chemical composition and $T_{\rm M}$ \cite{VanDenBerg1972,HorwoodJSolidStateChem1976}. This value is consistent with the $T_{\rm M}$ determined from the magnetic susceptibility measurements [Fig. \ref{figs2}(b)]. We find that the crystal structures at both 275 K and 100 K are well described by the NiAs-type structure within the space group $P6_3/mmc$. We note that if the symmetry were lowered to $P\bar{6}2c$ (a space group commonly reported for samples closer to ideal stoichiometry \cite{Keller-BesrestJSolidStateChem1990,BansalNP2020}), additional diffraction reflections would be expected at positions such as (1/3, 1/3, 0), however, no such reflections are resolved in our 100 K data [see Fig. \ref{figs3}(c)]. This indicates that, for this composition, either no structural distortion occurs or that any distortion is too small to be detected within the accuracy of our measurements. Nevertheless, slight deviation away from the NiAs-type structure can be caused by the iron deficiency, which is known and studied for years in h-FeS \cite{Nakazawa1971,WangPhaseTransition2005,ElliotActaCryst2010}.

\begin{figure}[h!]
	\hspace{0cm}\includegraphics[width=16cm]{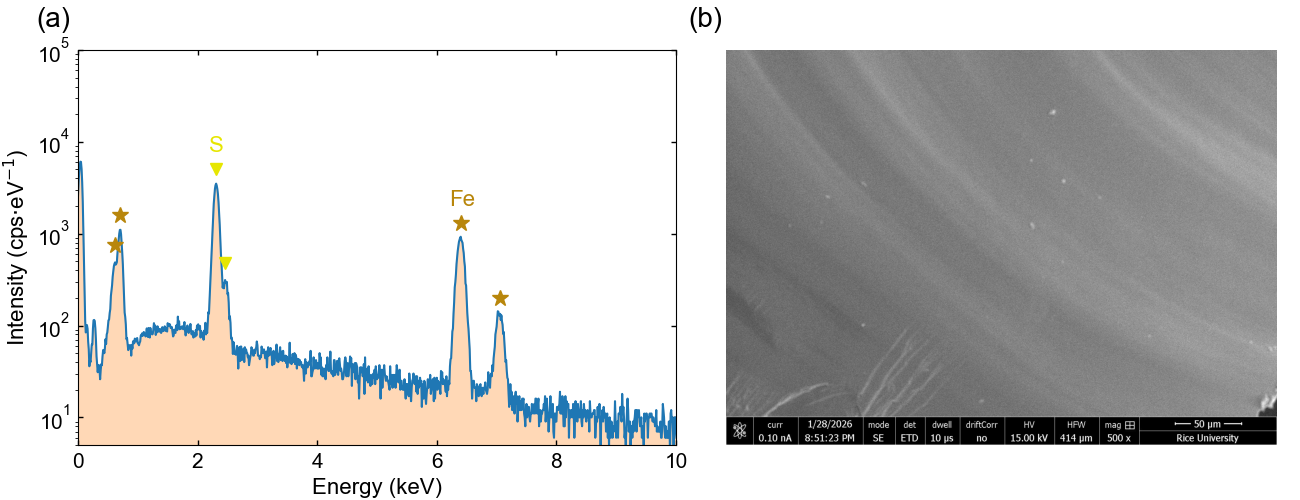}
	\caption{(a) Energy-dispersive X-ray spectroscopy acquired on the cleaved surface of h-FeS. The characteristic peaks of Fe and S are marked by stars and triangles, respectively. (b) Transmission electron microscopy image of the surface.}
	\label{figsx01}
\end{figure}

The chemical composition of exactly the same sample was further checked with energy-dispersive X-ray spectroscopy. Fig. \ref{figsx01}(a) shows one spectrum taken on a cleaved surface [Fig. \ref{figsx01}(b)]. The characteristic peaks of Fe and S can be well resolved. Elemental analysis indicates the ratio of Fe : S is close to 0.93 : 1 (see Table \ref{tb3}), which is consistent with the magnetic susceptibility and XRD results.

\section{Electrical transport measurement under uniaxial strain}

\begin{figure}[h!]
	\hspace{0cm}\includegraphics[width=13cm]{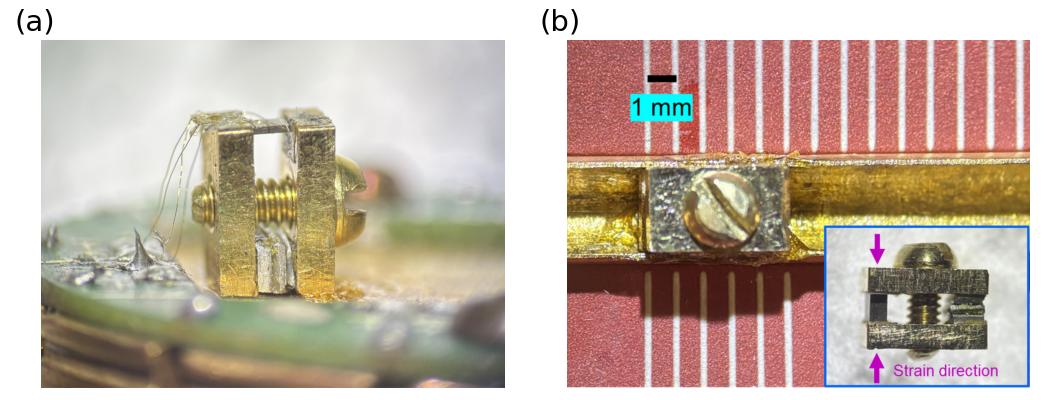}
	\caption{(a) Strain device with an h-FeS sample mounted on a standard PPMS resistivity puck. (b) Strain device with an h-FeS sample assembled on a standard VSM brass sample stick. The inset at the lower right shows a side view of the device, with the strain direction indicated.}
	\label{figs5}
\end{figure}

Electrical transport measurements were performed using a Quantum Design PPMS equipped with the electrical transport option. The h-FeS single crystals used in these measurements were cut and polished into approximate thin cuboids prior to making electrical contacts. Uniaxial compressive strain was applied using a small home-made mechanical strain device fabricated from brass [Fig. \ref{figs5}(a)]. The magnitude of the applied strain was estimated from the displacement associated with the advancement of the screw in the strain device and the Young’s modulus of h-FeS ($\sim$20 GPa), which was determined from the acoustic phonon dispersion [see Fig. 4(e) and (f) of the main text and also \cite{BansalNP2020}]. The Hall resistivity was extracted from the measured Hall signal by antisymmetrizing the data according to
\begin{equation}
	\rho_{yx} (B) = [ \rho (+ B ) -  \rho (- B)]/2,
\end{equation}
thereby eliminating contributions arising from voltage-probe misalignment.

\begin{figure}[h!]
	\hspace{0cm}\includegraphics[width=8cm]{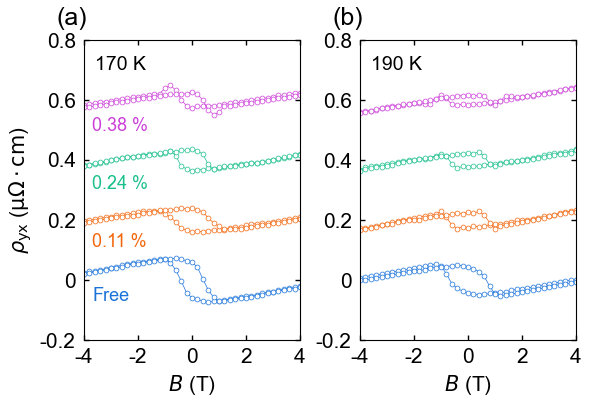}
	\caption{Magnetic-field dependence of the Hall resistivity measured at 170 K (a) and 190 K (b) under compressive strain applied along [$1\bar{1}0$]. The data are vertically offset by 0.2 $\mu \Omega\cdot\rm{cm}$ for clarity.}
	\label{figs6}
\end{figure}

Fig. \ref{figs6} shows the magnetic-field dependence of the Hall resistivity measured at 170 K and 190 K under compressive strain applied along [$1\bar{1}0$], which supplements the data presented in Fig. 2(a)-(d) of the main text.

We also measured the Hall resistivity on exactly the same h-FeS sample used for the magnetization measurements, with the data shown in Fig. 1(f) and Fig. 3 of the main text, as well as in Fig. \ref{figs7} and \ref{figs8} below. As presented in Fig. \ref{figsx03}, under $\sim$0.08 \% compressive strain applied along [$1\bar{1}0$], the magnitude of the spontaneous anomalous Hall effect (AHE) above $T_{\rm{M}}$ is suppressed. This behavior is consistent with the data shown in Fig. 2 of the main text, which were obtained from a separate sample. The Hall resistivity and magnetization measurements performed on exactly the same sample further consolidate the conclusion that both the spontaneous AHE and the tiny net magnetization can be simultaneously suppressed by the in-plane compressive strain.

\begin{figure}[h!]
	\hspace{0cm}\includegraphics[width=15cm]{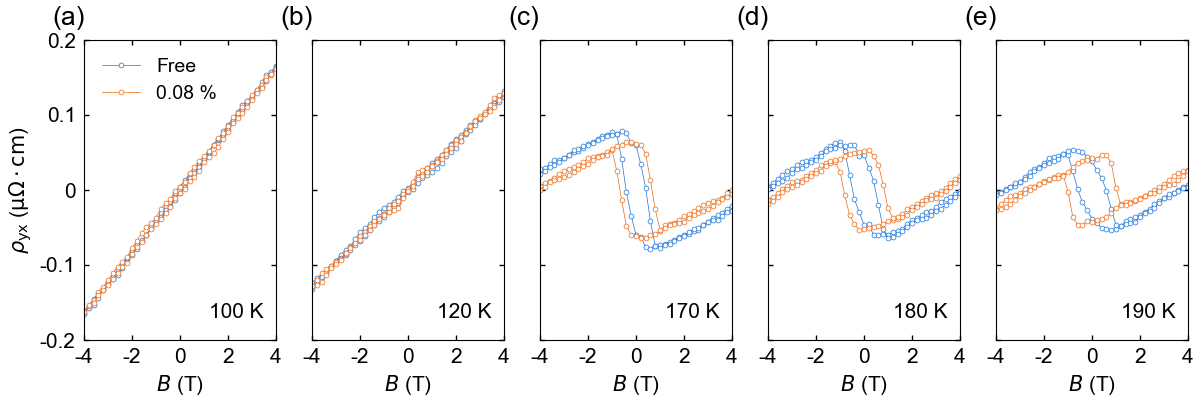}
	\caption{Magnetic-field dependence of the Hall resistivity measured at selected temperatures without strain and under $\sim$0.08 \% compressive strain applied along [$1\bar{1}0$].}
	\label{figsx03}
\end{figure}

\section{Magnetization measurement under uniaxial strain}

Magnetization measurements were also performed using a Quantum Design PPMS equipped with a VSM. Uniaxial compressive strain was applied using the same strain device as that employed in the electrical transport measurements [Fig. \ref{figs5}(b)], ensuring direct comparability between the two types of measurements. The magnitude of the applied strain was estimated using the same method as in the electrical transport experiments. The compressive strain was applied along [$1\bar{1}0$], and the magnetic field was applied along the $c$-axis.

\begin{figure}[h!]
	\hspace{0cm}\includegraphics[width=9cm]{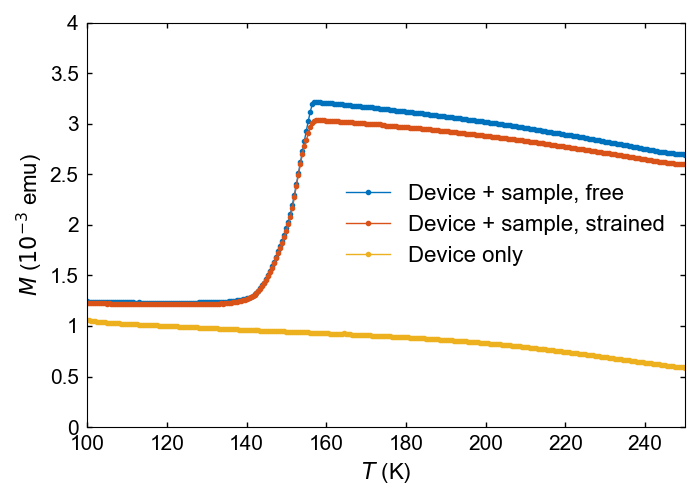}
	\caption{Temperature dependence of the magnetization measured at 1 T with the magnetic field applied along the $c$-axis. Data are shown for the strain device with an h-FeS sample in the free and strained states, together with the background signal from the device only.}
	\label{figs7}
\end{figure}

\begin{figure}[h!]
	\hspace{0cm}\includegraphics[width=12cm]{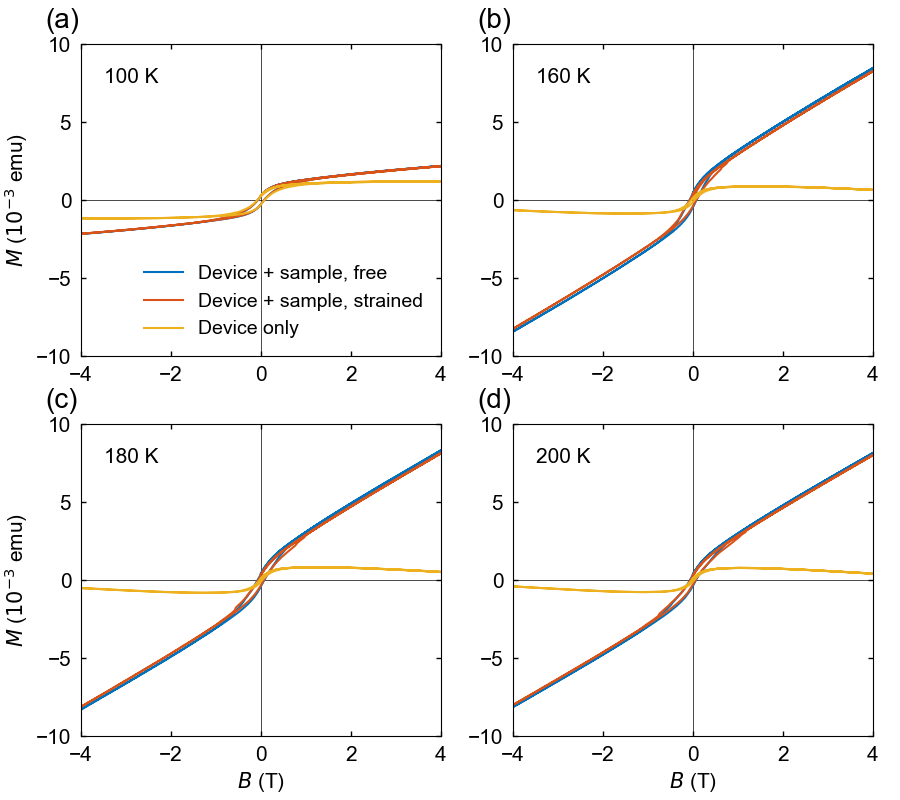}
	\caption{Isothermal magnetization along the $c$-axis at four temperatures for the strain device with an h-FeS sample in the free and strained states, together with the background signal from the device only.}
	\label{figs8}
\end{figure}

Since the mass of the strain device is much larger than that of the measured h-FeS sample ($\sim$8.7 mg), the background signal from the strain device cannot be neglected. Therefore, measurements of the strain device without the sample were also performed. These data were used as a background and subtracted from the measurements with the sample. Fig. \ref{figs7} and Fig. \ref{figs8} show the raw temperature- and field-dependent magnetization data, respectively, for the strain device with the sample in the free and strained states, as well as for the device only.

\begin{figure}[h!]
	\hspace{0cm}\includegraphics[width=18cm]{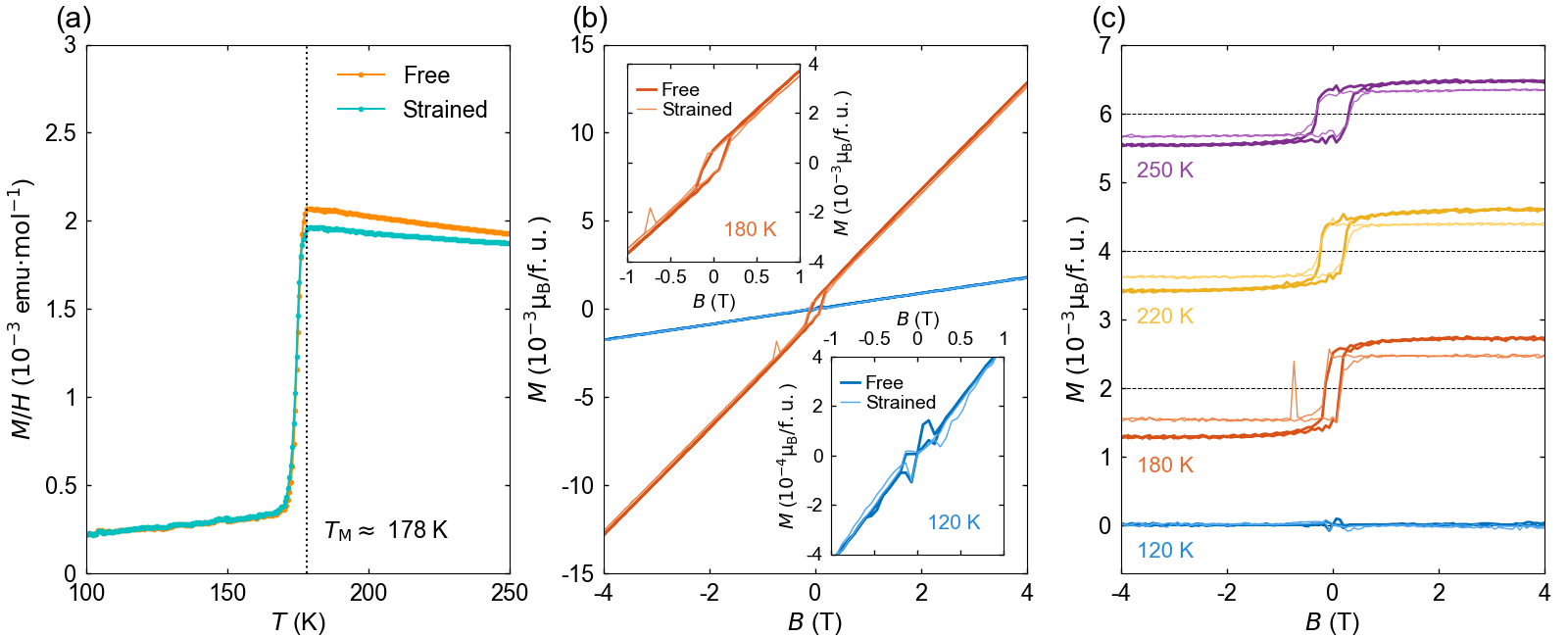}
	\caption{Temperature dependence of the magnetic susceptibility measured with and without compressive strain applied along [$1\bar{1}0$] under a magnetic field of 1 T along the $c$-axis. Dashed vertical line marks $T_{\rm{M}}$. (b) Isothermal magnetization along the $c$-axis with and without the strain at 120 K ($< T_{\rm{M}}$) and 180 K ($> T_{\rm{M}}$). Insets show the zoom-in view of data within $\pm$ 1 T. (c) Isothermal magnetization along the $c$-axis at selected temperatures after subtracting the high-fields linear part. Light and dark colors denote data with and without strain, respectively. Data are evenly offset 0.002 $\mu_{\rm{B}}$/f.u. for clarity.}
	\label{figsx02}
\end{figure}

To check the reproducibility of the strain effect on the magnetization, we have repeated the measurements on multiple samples. Fig. \ref{figsx02}(a) shows the temperature dependence of the magnetic susceptibility ($M/H$) on another h-FeS sample, which is measured under a 1 T magnetic field along the $c$-axis. We can see that $M/H$ is reduced by a compressive strain applied along [$1\bar{1}0$] above $T_{\rm{M}}$ ($\approx$ 178 K) but is almost unchanged below $T_{\rm{M}}$. The reduced $M/H$ is due to the suppressed spontaneous magnetization, as can be confirmed by the isothermal magnetization data shown in Fig. \ref{figsx02}(b) and (c). These results are consistent with the data shown in the main text [Fig. 1(f) and Fig. 3], although the $T_{\rm{M}}$s are different for the two samples.

\section{Neutron diffraction under uniaxial strain and magnetic fields}
\subsection{Neutron diffraction under uniaxial strain}
\begin{figure}[h!]
	\hspace{0cm}\includegraphics[width=17cm]{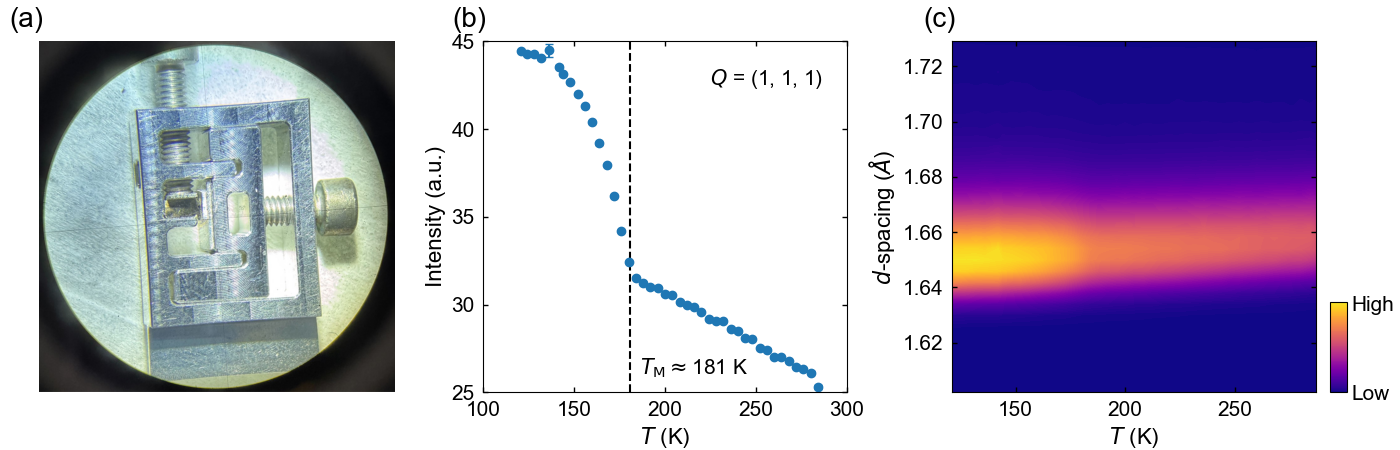}
	\caption{(a) Strain device with an h-FeS sample mounted for neutron diffraction measurements. (b) Temperature dependence of the diffraction intensity at the (1, 1, 1) Bragg peak. Dashed line marks the $T_{\rm M}$. (c) Contour plot for the temperature evolution of the diffraction intensity in the $d$-spacing region around the (1, 1, 1) Bragg peak.}
	\label{figs9}
\end{figure}

Neutron diffraction experiments were conducted on the CORELLI spectrometer installed at the Spallation Neutron Source of Oak Ridge National Laboratory \cite{YeJAC2018}. Uniaxial strain was applied using a home-made strain device fabricated from aluminum [Fig. \ref{figs9}(a)] \cite{LiuArxiv2025}. A piece of h-FeS single crystal with a mass of about 60 mg was used, and the strain was applied along [$1\bar{1}0$].The magnitude of the applied strain is approximately 100 MPa for the strained state, as estimated from the displacement associated with the advancement of the screw in the strain device and the Young’s modulus of h-FeS \cite{LiuArxiv2025}. Fig. \ref{figs9}(b) and (c) show the temperature dependence of the diffraction intensity at the (1, 1, 1) Bragg peak. A clear anomaly is observed at $T_{\rm M} \approx $ 181 K, consistent with a change in the spin configuration.

To get the information about magnetic domain population, the integrated peak intensity, which was obtained with Mantid \cite{Arnold2014}, was compared with the calculated intensity. For a magnetic Bragg peak, the integrated intensity is proportional to the modulus square of the magnetic structure factor $\textbf{F}_{M}(\textbf{Q})$ \cite{Shirane2002}
\begin{equation}
	I_{M}(\textbf{Q})=A|\textbf{F}_{M}(\textbf{Q})|^2,
\end{equation}
where
\begin{equation}
	\textbf{F}_{M}(\textbf{Q})=\sum_{j}\frac{\gamma r_0}{2} g_jf_j(Q)\textbf{S}_{\perp j}e^{i\textbf{Q}\cdot\textbf{r}_j}e^{-W_j}.
\end{equation}
$\textbf{S}_{\perp j}$ is the spin size at site $j$ that is detectable by neutrons
\begin{equation}
	\textbf{S}_{\perp j}=\textbf{S}_j-\hat{\textbf{Q}}(\hat{\textbf{Q}}\cdot\textbf{S}_j),
\end{equation}
where $\hat{\textbf{Q}}$ is the unit vector of $\textbf{Q}$ and $\textbf{S}_j$ is the spin vector at site $j$. $f_j(Q)$ and $e^{-W_j}$ are magnetic form factor and Debye-Waller factor, respectively. The magnetic moment at site $j$ in Bohr magneton is $g_jS_j$. The term $\frac{\gamma r_0}{2}$ in (3) containing the classical electron radius ($r_0$) and gyromagnetic ratio ($\gamma$) acts as an effective scattering length of per Bohr magneton, which is 2.695 fm. For h-FeS within the NiAs-type structure, there are two Fe ions occupying the 2$a$ Wyckoff site in each unit cell (Table \ref{tb1} and Table \ref{tb2}). The site positions are (0, 0, 0) and (0, 0, 0.5). We assume the spin vectors at these two positions are $\textbf{S}_1$ and $\textbf{S}_2$, respectively. If we ignore the tiny spin canting, $\textbf{S}_1$ and $\textbf{S}_2$ are opposite with each other
\begin{equation}
	\textbf{S}_1 = -\textbf{S}_2.
\end{equation}
For the six equivalent peaks shown in Fig. 4(a)-(d) of the main text, we can write the calculated diffraction intensity as
\begin{equation}
	I_{M}(\textbf{Q}) = A^{\prime}[\textbf{S}_1^2-(\hat{\textbf{Q}}\cdot\textbf{S}_1)^2],
\end{equation}
where $\textbf{Q}$ can be one of the wave vectors for Bragg peaks $(1, 1, 1)$, $(-1, 2, 1)$, $(-2, 1, 1)$, $(-1, -1, 1)$, $(1, -2, 1)$, and $(2, -1, 1)$. We have assumed the $g$-factor, form factor, and Debye-Waller factor are the same for these six peaks and have absorbed them into the new proportional coefficient $A^{\prime}$. By considering the three magnetic domains above $T_{\rm M}$, the total intensity can be written as
\begin{equation}
	I_{M}^{\rm{Total}}(\textbf{Q}) = a_{\rm{I}}[\textbf{S}_{\rm{I},1}^2-(\hat{\textbf{Q}}\cdot\textbf{S}_{\rm{I},1})^2]+a_{\rm{II}}[\textbf{S}_{\rm{II},1}^2-(\hat{\textbf{Q}}\cdot\textbf{S}_{\rm{II},1})^2]+a_{\rm{III}}[\textbf{S}_{\rm{III},1}^2-(\hat{\textbf{Q}}\cdot\textbf{S}_{\rm{III},1})^2],
\end{equation}
where $\textbf{S}_{\rm{I},1}$, $\textbf{S}_{\rm{II},1}$, and $\textbf{S}_{\rm{III},1}$ represent spin evctors at site position (0, 0, 0) for domains I, II, and III, respectively. $\textbf{S}_{\rm{I},1}$, $\textbf{S}_{\rm{II},1}$, and $\textbf{S}_{\rm{III},1}$ are conntected by three-fold rotational ($C_3$) symmetry. By comparing the integrated intensity from the experiment with the calculation with (7), we can get the coefficients $a_{\rm{I}}$, $a_{\rm{II}}$, and $a_{\rm{III}}$. The ratio $\alpha_{p}$ for magnetic domain $p$ ($p$ = I, II, and III) can be obtained as
\begin{equation}
	\alpha_{p} = \frac{a_{\rm{p}}}{a_{\rm{I}}+a_{\rm{II}}+a_{\rm{III}}}.
\end{equation}

Next, we consider the tiny spin canting, which results in a ferromagnetic component along the $c$-axis. If we assume the additional ferromagnetic spin component at every Fe ion is $\Delta S$, its contribution to the diffraction intensity can be written as
\begin{equation}
	I_{M}(\textbf{Q}) = B\Delta S^2[1-Q_{\rm{z}}^2/Q^2][2+2{\rm{cos}}\pi L],
\end{equation}
where 
\begin{equation}
	Q_{\rm{z}} = \frac{2\pi}{c}L
\end{equation}
and $B$ is another proportionality coefficient. The ferromagnetic moment is on the order of $10^{-3}$ $\mu_{\rm{B}}$, whereas the localized moment per Fe ion is approximately 3 $\mu_{\rm{B}}$ at 300 K \cite{BansalNP2020}. According to Eqs. (6) and (9), the ferromagnetic contribution is therefore only on the order of $10^{-7}$ of the antiferromagnetic contribution. On the other hand, the integrated intensities at (1, 0, 0) and (1, 1, 1) are 3128 a.u. and 1096 a.u., respectively, with corresponding uncertainties of 293 a.u. and 33 a.u. Consequently, the expected intensity change due to spin-canting variations is well below the experimental accuracy.

\subsection{Neutron diffraction under magnetic fields}

\begin{figure}[h!]
	\hspace{0cm}\includegraphics[width=16cm]{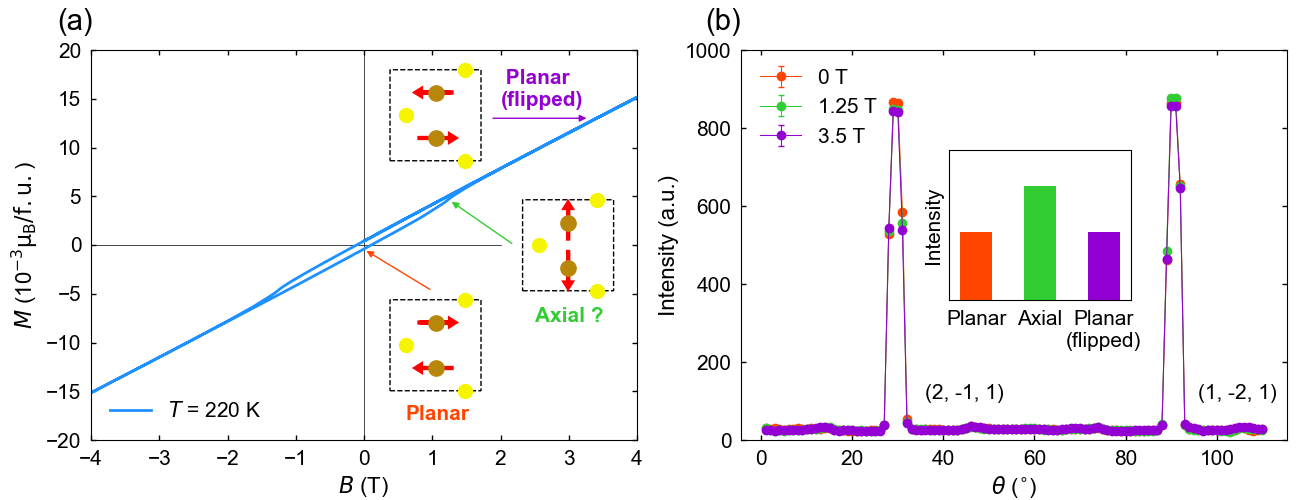}
	\caption{(a) Isothermal magnetization measured at 220 K with the magnetic field applied along the $c$-axis on the same h-FeS single crystal used for the neutron diffraction experiments. Insets schematically illustrate the possible spin configurations at 0 T, 1.25 T, and 3.5 T. (b) Angular ($\theta$) dependence of the diffraction intensity at a momentum-transfer magnitude corresponding to the (1, 1, 1) peak in the ($H$, $K$, 1) plane. The inset shows the calculated diffraction intensities for the spin configurations illustrated in (a).}
	\label{figs10}
\end{figure}

\begin{figure}[h!]
	\hspace{0cm}\includegraphics[width=17cm]{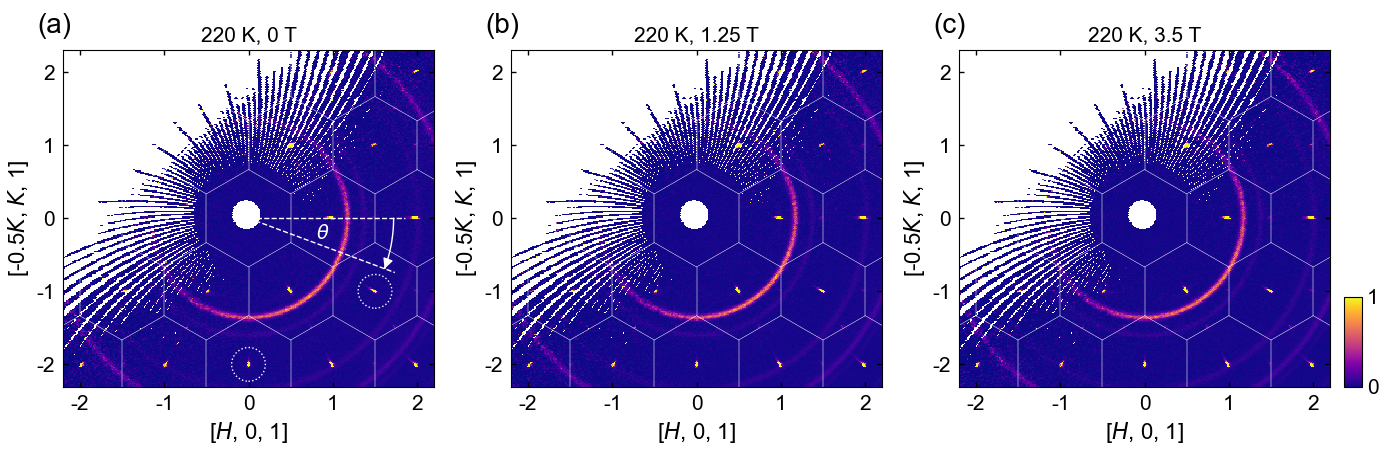}
	\caption{Neutron diffraction patterns of the ($H$, $K$, 1) plane measured at 220 K under magnetic fields of 0 T (a), 1.25 T (b), and 3.5 T (c). White hexagons denote the Brillouin zone boundaries. Dashed circles in (a) highlight the Bragg peaks at (2, $-1$, 1) and (1, $-2$, 1).}
	\label{figs11}
\end{figure}

We also conducted neutron diffraction measurements under magnetic fields applied along the $c$-axis (without strain) to explore the novel spin-flip transition proposed in \cite{TakagiNM2025}. In this experiment, we used an h-FeS single crystal with a mass of approximately 65 mg. The Mori transition temperature $T_{\rm M}$ of this sample is about 188 K, as determined from magnetic susceptibility measurements. The $c$-axis isothermal magnetization at 220 K is shown in Fig. \ref{figs10}(a). Our neutron diffraction measurements were also performed at 220 K and followed the same magnetization process. The diffraction patterns of the ($H$, $K$, 1) plane at 0 T (after returning from $-3.5$ T), 1.25 T, and 3.5 T are presented in Fig. \ref{figs11}. Owing to the magnet used in our experiment, the detectable out-of-plane range in reciprocal space is significantly limited, and only about half of the ($H$, $K$, 1) plane could be measured. No significant changes are observed in the diffraction patterns under these three conditions. This conclusion is further supported by the angular dependence of the diffraction intensity at a momentum-transfer magnitude corresponding to the (1, 1, 1) peak [Fig. \ref{figs10}(b)]. The definition of the angle ($\theta$) is illustrated in Fig. \ref{figs11}(a). As shown, the intensities of the (2, $-1$, 1) and (1, $-2$, 1) peaks remain essentially unchanged.

Next, we discuss the implications of these results on the underlying magnetic structures. At 0 T after coming back from $-3.5$ T, h-FeS exhibits a planar spin configuration [``Planar'' in the inset of Fig. \ref{figs10}(a)] with a small negative ferromagnetic moment. At 3.5 T ramped from this state, h-FeS also exhibits a planar spin configuration but with all the spins flipped [``Planar (flipped)'' in the inset of Fig. \ref{figs10}(a)] \cite{TakagiNM2025}. These two states are connected by time reversal symmetry. Although the directions of the spin canting are opposite, they show the same magnetic susceptibility and a basically identical diffraction pattern (as discussed above, the spin canting is beyond the experimental accuracy). This is consistent with the data shown in Fig. \ref{figs10}(b).

Based on the magnetization data shown in Fig. \ref{figs10}(a), h-FeS at 1.25 T is in an intermediate state during the transition from the `Planar'' state to the ``Planar (flipped)'' state. One possible scenario to achieve such a transition is a gradual rotation of all spins by 180$^\circ$ about the $c$-axis. However, we consider this scenario unlikely because the applied magnetic field is also along the $c$-axis. An alternative scenario is that all spins rotate by 180$^\circ$ about the $a$-axis, with 1.25 T corresponding to a field near the configuration where the spins point along the $c$-axis [the ``Axial'' state in the inset of Fig. \ref{figs10}(a)]. For this ``Axial'' state, the diffraction intensities at (2, $-1$, 1) and (1, $-2$, 1) are expected to be approximately 1.7 times larger than those of the ``Planar'' and Planar (flipped)'' states [see the inset of Fig. \ref{figs10}(b)]. Our experimental data at 1.25 T clearly contradict this expectation, and thus this scenario can also be ruled out. Therefore, a more likely scenario is that the transition from the ``Planar'' state to the ``Planar (flipped)'' state is first order. Any static experimental probe, such as the neutron diffraction measurements we performed, would be unable to detect the intermediate state.

\section{Inelastic neutron scattering measurement}

\subsection{Experimental setup and spin-wave gap analysis}
\begin{figure}[b!]
	\hspace{0cm}\includegraphics[width=15cm]{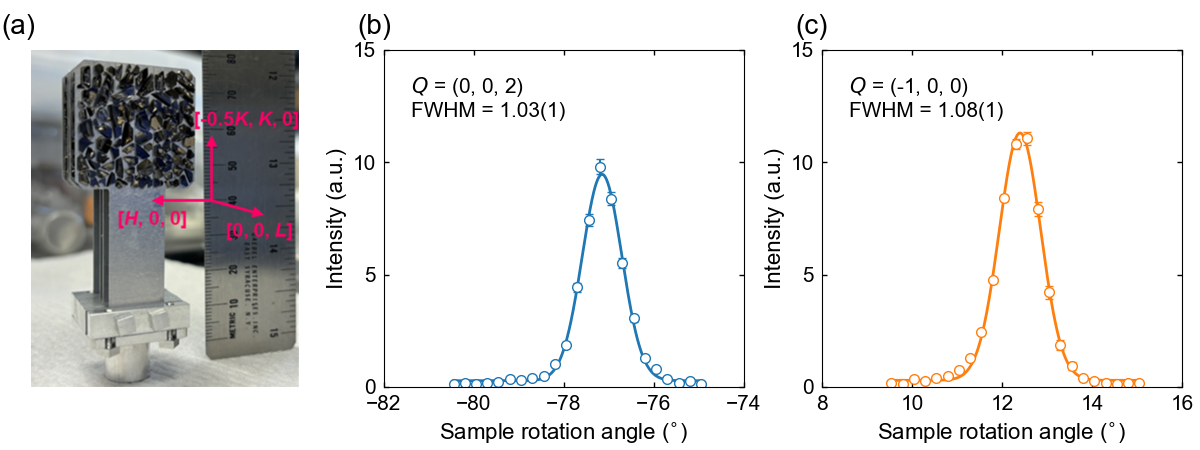}
	\caption{(a) Coaligned h-FeS single-crystal array. (b) and (c) Rocking scans performed around the structural Bragg peaks (0, 0, 2) and ($-1$, 0, 0), respectively. The solid curves represent fits using Gaussian profiles. FWHM denotes the full width at half maximum of the fitted Gaussian profile.}
	\label{figs12}
\end{figure}

Inelastic neutron scattering (INS) experiments were performed on the ARCS spectrometer at the Spallation Neutron Source, Oak Ridge National Laboratory \cite{Abernathy2012}. A coaligned h-FeS single-crystal array with a total mass of $\sim$19.5 g was used in this experiment [Fig. \ref{figs12}(a)]. The ($H$, 0, $L$) plane (within the NiAs-type structure) was aligned in the horizontal scattering plane [Fig. \ref{figs12}(a)]. The overall sample mosaic is approximately $1^\circ$, as determined from rocking scans of the structural Bragg peaks (0, 0, 2) and ($-1$, 0, 0) [Fig. \ref{figs12}(b) and (c)].

The $T_{\rm M}$ of the single-crystal array lies between 150 K and 200 K, as inferred from the temperature dependence of the (0, 0, 1) magnetic Bragg peak [inset of Fig. 4(g)]. This range approximately corresponds to $x = 0.92 \sim 0.94$ in Fe$_{x}$S \cite{HorwoodJSolidStateChem1976}. Data reduction and analysis were carried out using Mantid \cite{Arnold2014} and Horace \cite{Ewings2016}. The incident neutron energies ($E_i$s), chopper frequencies, energy resolutions at zero energy transfer, and momentum integration ranges used for the presented data are summarized in Table \ref{tb4}.

The spin-wave gap size below $T_{\rm M}$ is determined by fitting the data [Fig. 4(g) of the main text] with an error function multiplied by the Bose factor
\begin{equation}
	I (E, T)=\left[I_0+A\cdot \rm{erf}\left(\frac{\it{E}-\it{E}_{\rm{gap}}}{\sigma}\right)\right]\cdot \frac{1}{1-e^{-E/k_{\rm{B}}T}},
\end{equation}
where erf($x$) represents the error function, $I_0$, $A$, $\sigma$, and $E_{\rm{gap}}$ (the gap size) are fitting parameters.

\begin{figure}[t!]
	\hspace{0cm}\includegraphics[width=17cm]{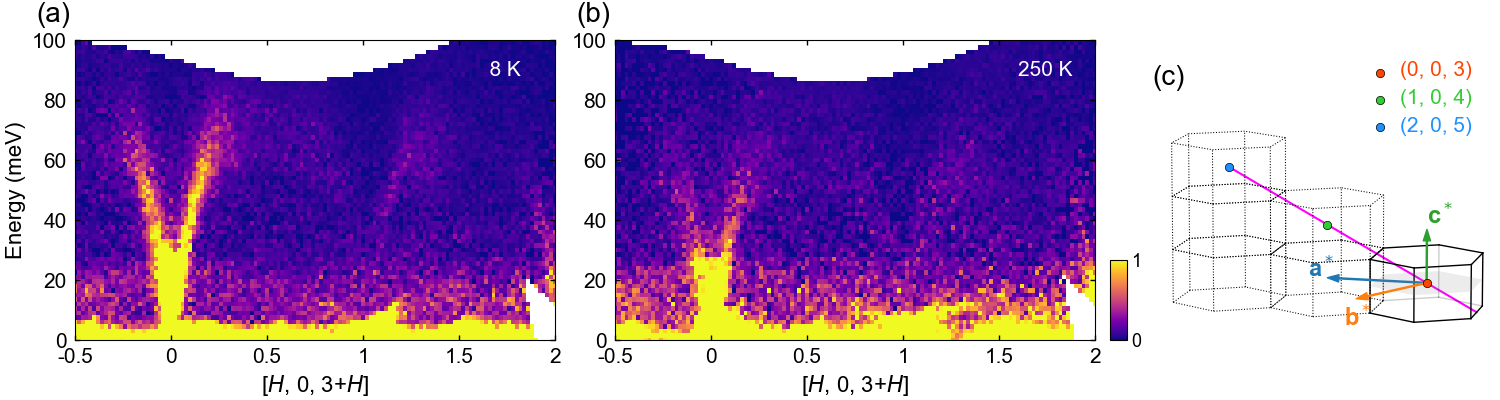}
	\caption{(a) and (b) Excitation spectra measured along the [$H$, 0, $3+H$] direction at 8 K and 250 K, respectively. (c) Schematic of multiple three-dimensional Brillouin zones of h-FeS. The magenta line indicates the momentum-space trajectory along which the excitation spectra in (a) and (b) are shown.}
	\label{figs13}
\end{figure}

\subsection{Chiral magnon splitting and high-energy spin waves}
Chiral magnon splitting is regarded as an important experimental observable for identifying altermagnets \cite{SmejkalPRL2023,GomonaynpjSpintronics2024,McClartyPRB2025}. It has recently been observed by INS in a series of altermagnet candidates, including $\alpha$-MnTe \cite{LiuPRL2024}, CrSb \cite{SinghArxiv2025}, Fe$_3$O$_4$ \cite{SunPRL2025}, MnF$_2$ \cite{FaureArxiv2025}, FeF$_2$ \cite{SearsArxiv2026}, and La$_2$O$_3$Mn$_2$Se$_2$ \cite{AsaiPRM2026}. We have also measured the high-energy spin waves in h-FeS, which enables us to examine whether chiral magnon splitting exists in this candidate material.

Fig. \ref{figs13}(a) presents the INS-measured excitation spectra along the [$H$, 0, $3+H$] direction at 8 K, well below $T_{\rm M}$. As illustrated in Fig. \ref{figs13}(c), this trajectory is where chiral magnon splitting is expected to occur in $g$-wave altermagnets \cite{KravchukPRB2025}. The spin wave emanating from (0, 0, 3) extends up to approximately 80 meV, with its intensity gradually decaying as the momentum moves away from (0, 0, 3). At 250 K, above $T_{\rm M}$, the spin wave shifts to slightly lower energies due to the closing of the anisotropy gap [see Fig. 4(e)-(h) of the main text]. Moreover, as h-FeS becomes more metallic at higher temperatures, the spin-wave excitations become further broadened.

We note that the spin configuration of h-FeS is similar to that of CrSb \cite{SinghArxiv2025} below $T_{\rm M}$ and to that of $\alpha$-MnTe \cite{LiuPRL2024} above $T_{\rm M}$. However, based on the data at hand shown in Fig. \ref{figs13}(a) and (b), we do not observe clear signatures of magnon splitting, in contrast to those reported in previously studied altermagnet candidates \cite{LiuPRL2024,SinghArxiv2025,SunPRL2025,FaureArxiv2025,SearsArxiv2026}. Further experiments with improved energy resolution and/or enhanced statistical quality may therefore be required.

\section{Magnetic symmetry and anomalous Hall effect}
In this section, we analyze the symmetry-allowed linear conductivities without an external magnetic field.
A conductivity tensor $\boldsymbol{\sigma}$ transforms invariantly under translation symmetry and covariantly under rotations. Therefore, it is constrained by the magnetic point group (MPG) of the system. More explicitly, the conductivity tensor satisfies the condition 
\begin{align}
	\boldsymbol{\sigma} = \frac{1}{|MPG|} \sum_{g\in MPG} g \, \boldsymbol{\sigma} \, g^{-1},
\end{align}
where $g$ represents an operator of the group element belonging to the MPG with an order of $|MPG|$. This allows us to identify symmetry-forbidden and symmetry-related entries in the tensor.

At $T>T_{\rm{N}}$, h-FeS is in the paramagnetic state with space group $P6_3/mmc$ and preserves time-reversal symmetry $\mathcal{T}$ \cite{TakagiNM2025}. In this phase, spontaneous AHE is forbidden due to $\mathcal{T}$ symmetry.

For $T_{\rm{M}} < T < T_{\rm{N}}$, the system develops an easy-plane antiferromagnetic (AFM) order \cite{TakagiNM2025}. There are three domains I, II, and III with their moments ordered along $[110]$, $[100]$, and $[010]$ directions, respectively [see Fig. 1(d) of the main text]. For a given single domain, the MPG is $2'm'm$, generated by $\{m_{001}|00\frac12 \}$, $\{2_{110}|000 \} {\mathcal T}$ and $\{m_{1\bar10}|00\frac12 \} {\mathcal T}$ symmetries. It constrains the conductivity tensor as 
\begin{equation}
	\boldsymbol{\sigma}=\begin{pmatrix}
		\sigma_{xx} & \sigma_{xy} &0 \\ 
		-\sigma_{xy} & \sigma_{yy} & 0 \\ 
		0 &0 & \sigma_{zz} 
	\end{pmatrix},
\end{equation}
where the $x$ axis is defined as the N\'eel vector direction of the AFM order. When in-plane strain is applied, domain I is disfavored and the symmetry is reduced to a subgroup containing a $z$-normal mirror $\{m_{001}|00\frac12 \}$. This is because strain breaks $\{2_{110}|000 \} {\mathcal T}$ and $\{m_{1\bar10}|00\frac12 \} {\mathcal T}$ symmetries. This symmetry still allows the same form of the conductivity tensor, in particular a finite $\sigma_{xy}$.

At $T <T_{\rm{M}}$, a Morin transition occurs, and the system transfers to a distinct easy-axis AFM order with the magnetic moments aligning along the $c$-axis \cite{TakagiNM2025}. The MPG is $\bar 6' 2m'$, where some key symmetries are $\{C_{3,001}|000\}$, $\{C_{2,110}|000\}$, $\{m_{001}|00\frac12 \}\mathcal{T}$, $\{m_{210}|00\frac12 \}\mathcal{T}$, and $\{m_{1\bar{1}0}|00\frac12 \}\mathcal{T}$. This MPG constrains the conductivity tensor as 
\begin{equation}
	\boldsymbol{\sigma}=\begin{pmatrix}
		\sigma_{xx} & 0 &0 \\ 
		0 & \sigma_{yy} & 0 \\ 
		0 &0 & \sigma_{zz} 
	\end{pmatrix}.
\end{equation}
When in-plane strain is applied along $[110]$, 
$\{C_{2,110}|000\}$, $\{m_{001}|00\frac12 \}\mathcal{T}$ and $\{m_{1\bar{1}0}|00\frac12 \}\mathcal{T}$ are preserved while other symmetries are broken. The corresponding MPG is reduced to $2m'm'$, and it constrains the conductivity tensor as
\begin{equation}
	\boldsymbol{\sigma}=\begin{pmatrix}
		\sigma_{xx} &0 &0 \\ 
		0 & \sigma_{yy} & \sigma_{yz}  \\ 
		0 &-\sigma_{yz} & \sigma_{zz} 
	\end{pmatrix}.
\end{equation}
When in-plane strain is applied along $[1\bar10]$, $\{C_{2,110}|000\}$, $\{m_{001}|00\frac12 \}\mathcal{T}$ and $\{m_{210}|00\frac12 \}\mathcal{T}$ are preserved while other symmetries are broken. The corresponding MPG is reduced to $m'2m'$, and it constrains the conductivity tensor as
\begin{equation}
	\boldsymbol{\sigma}=\begin{pmatrix}
		\sigma_{xx} &0 &\sigma_{xz} \\ 
		0 & \sigma_{yy} & 0  \\ 
		-\sigma_{xz} &0 & \sigma_{zz} 
	\end{pmatrix}.
\end{equation}
Here, the strain direction is defined as the $x$ axis, and the MPG symbols are given in the $xyz$ order. In conclusion, the applied strain does not generate an in-plane spontaneous AHE. Further analysis shows that an in-plane second-order nonlinear AHE may appear under applied strain, as permitted by the reduced magnetic symmetry.

\section{On the magnetocrystalline anisotropy of h-FeS}

The spin configuration of h-FeS is governed by magnetocrystalline anisotropy, which has been extensively studied since the 1960s \cite{AdachiJPSJ1961,Adachi1963,SatoJPSJ1966,AdachiJApplPhys1968,HorwoodJSolidStateChem1976}. In the following, we adopt the key results of these theories and refer the reader to \cite{AdachiJPSJ1961,Adachi1963,SatoJPSJ1966,AdachiJApplPhys1968,HorwoodJSolidStateChem1976} for full details.

Within the NiAs-type structure, the anisotropic free energy ($\mathcal{F}$) of the Fe ions can be expressed as a function of the moment canting angle ($\vartheta$) measured from the $c$-axis
\begin{equation}
	\mathcal{F}(\vartheta)
	= -c_1 k_{\rm{B}} T
	\ln \!\left\{
	\cosh \!\left[
	\frac{2(\it{\lambda} l_{\rm{z}} \cos\vartheta + \it{R})}{k_{\rm{B}} T}
	\right]
	\right\}
	+ \it{c}_{\rm{2}} K \cos^{\rm{2}} \vartheta.
\end{equation}
where the first term represents the anisotropy energy originating from the degenerate orbital state with a preferred orientation along the $c$-axis. This degeneracy is lifted by a weaker orthorhombic crystal field (arising from iron vacancies), resulting in an energy splitting of width $2R$. The second term describes the anisotropy energy favoring an easy axis within the $ab$-plane, which originates from perturbations of the nondegenerate orbital state as well as the magnetic dipolar interaction. Here, $c_1$ and $c_2$ are positive numerical coefficients for the respective terms. $\lambda$ denotes the strength of the spin-orbit coupling (SOC), and $l_{\rm{z}}$ is the $z$-component of the orbital angular momentum, with the $z$ axis chosen along the crystallographic $c$-axis of the NiAs-type structure. Because the orbital angular momentum of Fe$^{2+}$ ions with an electronic configuration [Ar]3$d^6$ is partially quenched, $l_{\rm{z}}$ effectively quantifies the residual orbital contribution, and the product $\lambda l_{\rm{z}}$ can be regarded as an effective SOC. For $T > T_{\rm M}$, the anisotropy constant $K$ is positive, ensuring that the spins favor an easy-plane configuration with $\vartheta$ close to $90^\circ$.

The spin direction can be derived by setting $\frac{\partial\mathcal{F}}{\partial\vartheta} = 0$, which gives
\begin{equation}
	\frac{c_2 K}{c_1 \lambda l_{\rm{z}}} \cos\vartheta
	=
	\tanh \!\left[
	\frac{2(\it{\lambda} l_{\rm{z}} \cos\vartheta + \it{R})}{k_{\rm{B}} T}
	\right] .
\end{equation}
By taking the high temperature limit and defining a canting angle ($\delta = 90^\circ - \vartheta$) measured from the $ab$-plane, we get 
\begin{equation}
	\sin\delta
	\approx
	\frac{2 \it{R}}
	{\dfrac{\it{c}_{\rm{2}} K k_{\rm{B}} T}
		{\it{c}_{\rm{1}} \lambda \it{l}_{\rm{z}}}
		- 2 \lambda \it{l}_{\rm{z}}} .
\end{equation}
Since $\delta$ is a small positive angle close to zero, this expression indicates that $\delta$ decreases with decreasing $\it{l}_{\rm{z}}$. Therefore, when the angular momentum is further quenched by the reduced symmetry of the crystal field (i.e., the $C_3$ is removed by the in-plane strain in our case), the canting angle is correspondingly suppressed. This behavior is consistent with our experimental observation that uniaxial compressive strain applied in the $ab$-plane reduces the spin canting. In addition, we note that the canting angle $\delta$ also decreases with increasing temperature, which naturally accounts for the reduced spin canting observed at elevated temperatures.

Therefore, we conclude that the established magnetocrystalline anisotropy theory can qualitatively interpret our experimental observations quite well. We hope that our study will motivate further theoretical investigations along this direction for h-FeS, as well as for other altermagnet candidates with the NiAs-type crystal structure.

\begin{table}[htbp]
	\caption{Results of single crystal XRD refinement at 275 K.}
	\centering
	\begin{tabularx}{0.8\textwidth}
		{l *{5}{>{\centering\arraybackslash}X}}
		\hline\hline
		\multicolumn{6}{c}{FeS, $P$6$_3/mmc$} \\
		\hline
		Radiation            & \multicolumn{5}{c}{Mo $K_{\alpha 1}$, $\lambda = 0.71073$ \AA} \\
		Temperature          & \multicolumn{5}{c}{274.9(2) K} \\
		Total reflections    & \multicolumn{5}{c}{1284} \\
		Unique reflections   & \multicolumn{5}{c}{75 ($I > 3\sigma$)} \\
		Extinction           & \multicolumn{5}{c}{0.066(5)} \\
		Refinement quality   & \multicolumn{5}{c}{$\chi^2 = 0.62$, $R_{\rm{p}} = 1.47$, $R_{\rm{wp}} = 2.33$} \\
		Lattice parameters   & \multicolumn{5}{c}{\makecell{$a=b=3.4519(1)$ \AA, $c=5.7619(3)$ \AA \\ $\alpha=\beta=90^\circ, \gamma=120^\circ$}} \\
		\midrule
		Atom & $x$ & $y$ & $z$ & Occ. & Wyckoff \\
		\midrule
		Fe   & 0   & 0   & 0   & 0.924(7) & 2$a$\\
		S    & 1/3 & 2/3 & 3/4 & 1        & 2$d$\\
		\midrule
		\multicolumn{6}{c}{Anisotropic displacement parameters (\AA$^{2}$)}\\
		
		\midrule
		Atom & \multicolumn{2}{c}{$U_{11} = U_{22} = 2U_{12} $} & $U_{33}$  & $U_{13} = U_{23}$ & $U_{\rm{eq}}$ \\
		\midrule
		Fe   & \multicolumn{2}{c}{0.0351(6)} & 0.0149(6)   & 0 & 0.0284(3)\\
		S    & \multicolumn{2}{c}{0.0097(6)} & 0.0198(8)   & 0 & 0.0131(4)\\
		\hline\hline
	\end{tabularx}
	\label{tb1}
\end{table}

\begin{table}[htbp]
	\caption{Results of single crystal XRD refinement at 100 K.}
	\centering
	\begin{tabularx}{0.8\textwidth}
		{l *{5}{>{\centering\arraybackslash}X}}
		\hline\hline
		\multicolumn{6}{c}{FeS, $P$6$_3/mmc$} \\
		\hline
		Radiation            & \multicolumn{5}{c}{Mo $K_{\alpha 1}$, $\lambda = 0.71073$ \AA} \\
		Temperature          & \multicolumn{5}{c}{100.0(1) K} \\
		Total reflections    & \multicolumn{5}{c}{1109} \\
		Unique reflections   & \multicolumn{5}{c}{69 ($I > 3\sigma$)} \\
		Extinction           & \multicolumn{5}{c}{0.057(27)} \\
		Refinement quality   & \multicolumn{5}{c}{$\chi^2 = 1.15$, $R_{\rm{p}} = 2.54$, $R_{\rm{wp}} = 5.18$} \\
		Lattice parameters   & \multicolumn{5}{c}{\makecell{$a=b=3.4380(2)$ \AA, $c=5.7624(5)$ \AA \\ $\alpha=\beta=90^\circ, \gamma=120^\circ$}} \\
		\midrule
		Atom & $x$ & $y$ & $z$ & Occ. & Wyckoff \\
		\midrule
		Fe   & 0   & 0   & 0   & 0.945(32) & 2$a$\\
		S    & 1/3 & 2/3 & 3/4 & 1        & 2$d$\\
		\midrule
		\multicolumn{6}{c}{Anisotropic displacement parameters (\AA$^{2}$)}\\
		
		\midrule
		Atom & \multicolumn{2}{c}{$U_{11} = U_{22} = 2U_{12} $} & $U_{33}$  & $U_{13} = U_{23}$ & $U_{\rm{eq}}$ \\
		\midrule
		Fe   & \multicolumn{2}{c}{0.0350(25)} & 0.0083(26)   & 0 & 0.0261(14)\\
		S    & \multicolumn{2}{c}{0.0043(24)} & 0.0158(32)   & 0 & 0.0081(15)\\
		\hline\hline
	\end{tabularx}
	\label{tb2}
\end{table}

\begin{table}[h]
	\caption{Elemental analysis report obtained from the spectrum shown in Fig. \ref{figsx01}(a).}
	\begin{ruledtabular}
		\begin{tabular}{cccc}
			Element&Weight percentage (\%)&Weight percentage error (\%)&Atomic percentage (\%)\\
			\midrule
			S&38.16&0.48&51.74\\
			Fe&61.84&0.97&48.26\\
			\midrule
			Total&100&-&100\\
		\end{tabular}	
	\end{ruledtabular}
	\label{tb3}
\end{table}

\begin{table}[t]
	\caption{Momentum-integration ranges, chopper frequencies, energy resolutions, and $E_i$s for the INS data presented in figures.}
	\begin{ruledtabular}
		\begin{tabular}{ccccc}
			Data &\makecell{Momentum-integration\\range (r.l.u.)}& \makecell{Chopper\\frequency (Hz)}&\makecell{Energy \\resolution (meV)}&$E_i$ (meV) \\
			\midrule
			Fig. 4(e) and (f) &\makecell{[$-0.05$, 0.05] along [$H$, 0, 0]\\ $$[$-0.05$, 0.05] along [$-K$, $0.5K$, 0]}& \makecell{120 (Fermi1)\\120 (Fermi2)}& 4 &60\\
			\hline
			Fig. 4(g) &\makecell{[$-0.05$, 0.05] along [$H$, 0, 0]\\ $$[$-0.05$, 0.05] along [$-K$, $0.5K$, 0]\\$$[$-0.05$, 0.05] along [0, 0, $L$]}& \makecell{120 (Fermi1)\\120 (Fermi2)}& 4 &60\\
			\hline
			Fig. 4(h) left &\makecell{[$-0.03$, 0.03] along [$H$, 0, 0]\\ $$[$-0.03$, 0.03] along [$-K$, $0.5K$, 0]}& \makecell{120 (Fermi1)\\240 (Fermi2)}& 0.5 &15\\
			\hline
			Fig. 4(h) right &\makecell{[$-0.03$, 0.03] along [$H$, 0, 0]\\ $$[$-0.03$, 0.03] along [$-K$, $0.5K$, 0]\\$$[$-0.03$, 0.03] along [0, 0, $L$]}& \makecell{120 (Fermi1)\\240 (Fermi2)}& 0.5 &15\\
			\hline
			Fig. \ref{figs13}(a) and (b) &\makecell{[$-0.05$, 0.05] along [$-K$, $0.5K$, 0]\\$$[$-0.05$, 0.05] along [$-0.2688L$, 0, $3+L$]}& \makecell{120 (Fermi1)\\ 600 (Fermi2)}& 5 &150\\
		\end{tabular}	
	\end{ruledtabular}
	\label{tb4}
\end{table}

\clearpage

\end{document}